\documentclass[usenatbib]{mn2e} 
\usepackage{epsfig}
\usepackage{rotating}

\newif\ifAMStwofonts
\AMStwofontstrue

\def\be{\begin{equation}}
\def\ee{\end{equation}}

\def\gtsima{$\; \buildrel > \over \sim \;$}
\def\ltsima{$\; \buildrel < \over \sim \;$}
\def\prosima{$\; \buildrel \propto \over \sim \;$}
\def\gsim{\lower.5ex\hbox{\gtsima}}
\def\lsim{\lower.5ex\hbox{\ltsima}}
\def\simgt{\lower.5ex\hbox{\gtsima}}
\def\simlt{\lower.5ex\hbox{\ltsima}}
\def\simpr{\lower.5ex\hbox{\prosima}}
\def\la{\lsim}
\def\ga{\gsim}

\def\Lya{Ly$\alpha$~}

\def\HI{\hbox{H$\,\rm \scriptstyle I\ $}}
\def\HII{\hbox{H$\,\rm \scriptstyle II\ $}} 
\def\HeI{\hbox{He$\,\rm \scriptstyle I\ $}}
\def\HeII{\hbox{He$\,\rm \scriptstyle II\ $}}
\def\HeIII{\hbox{He$\,\rm \scriptstyle III\ $}}

\title[The effect of helium on  hydrogen reionisation]{The effect of
  intergalactic helium on hydrogen reionisation: implications for the
  sources of ionising photons at \boldmath{$z>6$}}

\author[B. Ciardi et al.]
{B. Ciardi$^{1}$\thanks{E-mail:ciardi@mpa-garching.mpg.de}, J.S. Bolton$^{2}$, A. Maselli$^3$ and L. Graziani$^{1}$\\
$^1$ Max-Planck-Institut fuer Astrophysik,
Karl-Schwarzschild-Strasse 1, D-85748 Garching b. Muenchen, Germany\\
$^2$ School of Physics, University of Melbourne, Parkville, Victoria 3010, Australia\\
$^3$ EVENT Lab for Neuroscience and Technology, Universitat de Barcelona, Passeig de la Vall d'Hebron 171, 08035 Barcelona, Spain
}

\date{December 20 2011}    

\pagerange{\pageref{firstpage}--\pageref{lastpage}}
\pubyear{2011}

\begin{document}

\maketitle
\label{firstpage}

\begin{abstract}
We investigate the effect of primordial helium on hydrogen
reionisation using a hydrodynamical simulation combined with the
cosmological radiative transfer code {\tt CRASH}.  The radiative
transfer simulations are performed in a $35.12h^{-1}$ comoving Mpc box
using a variety of assumptions for the amplitude and power-law
extreme-UV (EUV) spectral index of the ionising emissivity at $z>6$.
We use an empirically motivated prescription for ionising sources
which, by design, ensures all of the models are consistent with
constraints on the Thomson scattering optical depth and the
metagalactic hydrogen photo-ionisation rate at $z \sim 6$.  The
inclusion of helium slightly delays reionisation due to the small
number of ionising photons which reionise neutral helium instead of
hydrogen.  However, helium has a significant impact on the thermal
state of the IGM during hydrogen reionisation.  Models with a soft EUV
spectral index, $\alpha=3$, produce IGM temperatures at the mean
density at $z\sim 6$, $T_{0}\simeq 10500\rm\,K$, which are $\sim 20$
per cent higher compared to models in which helium photo-heating is
excluded.  Harder EUV indices produce even larger IGM temperature
boosts by the end of hydrogen reionisation.  A comparison of these
simulations to recent observational estimates of the IGM temperature
at $z\sim 5$--$6$ suggests that hydrogen reionisation was primarily
driven by population-II stellar sources with a soft EUV index,
$\alpha\la 3$.  We also find that faint, as yet undetected galaxies,
characterised by a luminosity function with a steepening faint-end
slope ($\alpha_{\rm LF} \leq -2$) and an increasing Lyman continuum
escape fraction ($f_{\rm esc}\sim 0.5$), are required to reproduce the
ionising emissivity used in our simulations at $z>6$.  Finally, we
note there is some tension between recent observational constraints
which indicate the IGM is $>10$ per cent neutral by volume $z\sim 7$,
and estimates of the ionising emissivity at $z=6$ which indicate only
$1$--$3$ ionising photons are emitted per hydrogen atom over a Hubble
time at $z=6$.  This tension may be alleviated by either a lower
neutral fraction at $z\sim 7$ or an IGM which still remains a few per
cent neutral by volume at $z=6$.

\end{abstract}



\section{Introduction} \label{sec:intro}

The last decade has witnessed the establishment of the two key pieces
of observational evidence which presently shape our empirical
understanding of the hydrogen reionisation epoch.  The first is the
Thomson scattering optical depth inferred from observations of the
cosmic microwave background (CMB).  This measurement provides a
constraint on the integrated reionisation history, and is consistent
with hydrogen reionisation beginning no later than $z=10.6\pm 1.2$
\citep{Komatsu11}.  The second is the signature of \HI \Lya absorption
in the spectra of high redshift quasars; observations of the
\cite{GunnPeterson65} trough indicate that the intergalactic medium
(IGM) is largely ionised by redshifts less than $z \simeq 6$
(\citealt{Becker01,Fan06}).  These observational data therefore
broadly constrain hydrogen reionisation to the redshift range $z\simeq
6-12$.

Despite this progress, however, a detailed determination of the timing
and extent of hydrogen reionisation, as well as the exact nature of
the sources responsible for driving this process, remains elusive.
Due to the integral nature of the CMB constraint a wide range of
extended reionisation histories are compatible with the Thomson
scattering optical depth measurement.  The small neutral hydrogen
fractions, $x_{\rm HI}\sim 10^{-4}$, at which \Lya absorption
saturates also leave room for alternative interpretations of the
quasar data (e.g. \citealt{Songaila04,Becker07}).  Furthermore,
\cite{Mesinger10} has recently pointed out that even at $z\sim
5$--$6$, the IGM may still harbour large patches of neutral hydrogen;
the number of currently known quasar sight-lines is insufficient to
fully rule out this possibility with intergalactic \Lya absorption
observations alone.

One route to making further progress is therefore developing detailed
simulations
(e.g. \citealt{Ciardi.Ferrara.White_2003,Iliev.Mellema.Shapiro.Pen_2007,Trac.Cen_2007,Finlator09b,Aubert10,Baek_etal_2010}),
and semi-numerical/analytical models
(e.g. \citealt{Choudhury.Ferrara_2006,Zahn_etal_2007,Mesinger07,
  Santos_etal_2010,Shull12,Raskutti12}) which can be compared to these
data to make inferences about the reionisation process.  However, most
existing numerical simulations do not explore the effect of hard,
helium ionising photons on the thermal state of the IGM during
hydrogen reionisation (although see
e.g. \citealt{Sokasian2002,Paschos.Norman.Bordner.Harkness_2008,McQuinn2009}
for treatments of \HeII reionisation at $z\simeq 3$).  This renders
the comparison of these models to measurements of the IGM temperature
at $z<6$ problematic.  In addition, many numerical models
significantly over-predict the number of ionising photons in the IGM
relative to observational constraints on the \HI photo-ionisation rate
from the \Lya forest at $z\sim 5$--$6$.  These data are consistent
with $\sim 1$--$3$ ionising photons emitted per hydrogen atom over a
Hubble time at $z=6$.  As a result, in order for hydrogen reionisation
to complete by $z=6$ {\it and} simultaneously match observational
constraints from the CMB and the background photo-ionisation rate at
$z\leq 6$, reionisation must be an extended process where the ionising
emissivity increases at $z>6$
(\citealt{MiraldaEscude2003,Meiksin05,BoltonHaehnelt07,HaardtMadau2011,McQuinn11}).
Correctly matching these ``post-reionisation'' constraints therefore
has important implications for reionisation and the properties of the
ionising sources in the early Universe.

In this work we address these issues using radiative transfer
simulations of reionisation performed using the code {\tt CRASH}
\citep{Ciardi_etal_2001,Maselli.Ferrara.Ciardi_2003,Maselli.Ciardi.Kanekar_2009,
  Partl_etal_2011}.  {\tt CRASH} is a 3D Monte Carlo based code which
follows the propagation of ionising photons (from both point sources
and diffuse radiation) and self-consistently calculates the evolution
of the gas temperature and ionisation state of hydrogen and helium in
the IGM.  Our approach differs from previous studies in two important
ways.  Firstly, we include the effect of helium ionising photons on
the progression of hydrogen reionisation.  This is especially
important for computing the thermal state of the IGM
(e.g. \citealt{Tittley07,Cantalupo11,PawlikSchaye11}), and it enables
us to directly compare our simulations to recent measurements of the
IGM temperature at $z=5$--$6$ (\citealt{Becker11,Bolton12}).
Secondly, instead of using a numerical sub-grid model for the sources
of ionising photons, the ionising emissivity in our simulations is
matched to the CMB and \Lya forest observational constraints by
design.  The goal of this empirical approach is to explore the
consequences of satisfying these observational constraints for
reionisation models from the outset, instead of tuning free parameters
and/or sub-grid prescriptions within the simulations.

The structure of this paper is as follows.  We begin in Section 2 with
a discussion of the empirically motivated reionisation models used in
our analysis, and continue in Section 3 with a description of our
numerical simulations.  In Section 4 we demonstrate that our RT
simulations match the observational constraints on the Thomson
scattering optical depth and photo-ionisation rate inferred from the
\Lya forest at $z\sim 6$, before going on to discuss in detail the
effect of including helium on the ionisation and thermal state of the
IGM in Section 5.  We perform a comparison of our simulations to the
observational data in Section 6 and discuss the implications for the
the properties of ionising sources at $z>6$.  Finally, we summarise
and conclude in Section 7.  An appendix
presenting selected numerical convergence tests of our simulations is
provided at the end of the paper.
Throughout the paper, the following
cosmological parameters are used: $\Omega_{\Lambda}=0.74$,
$\Omega_m=0.26$, $\Omega_b=0.024 h^2$, $h=0.72$, $n_s=0.95$ and
$\sigma_8=0.85$, where the symbols have the usual meaning.


\section{The reionisation history}\label{analytic}

The primary goal of this work is to model the effect of hydrogen and
helium ionising photons on the IGM, rather than self-consistently
modelling star formation and feedback effects during reionisation.
Rather than use a sub-grid prescription for modelling the production
of ionising photons in our simulations, we shall instead adopt an
empirically motivated approach which satisfies the observational
constraints from the CMB and \Lya forest at $z\sim 6$ by design.  We
achieve this by using a simple semi-analytical model to initially
guide the choice of ionising emissivity within our radiative transfer
simulations.

We first define the total comoving hydrogen ionising emissivity to be
$\epsilon_{\rm HI}\rm\,[s^{-1}\,Mpc^{-3}]$, where the volume filling
factor of \HII is obtained by solving (e.g. \citealt{Madau99})

\begin{equation} 
\frac{dQ_{\rm HII}}{dt} = \frac{ \epsilon_{\rm HI}}{\langle n_{\rm H} \rangle} - Q_{\rm HII}C_{\rm HII}\frac{\langle n_{\rm e} \rangle_{\rm HII}}{a^{3}}\alpha_{\rm HII}(T). 
\label{eq:ffH}
\end{equation}

\noindent
Here $\alpha_{\rm HII}(T)$ is the case-A recombination coefficient,
$\langle n_{\rm H} \rangle$ is the mean comoving hydrogen number
density, $\langle n_{\rm e} \rangle$ is the mean comoving electron
number density, $a=(1+z)^{-1}$ and $C_{\rm HII}=\langle n_{\rm
  HII}^{2} \rangle/ \langle n_{\rm HII} \rangle^{2}$ is the clumping
factor of hydrogen within the ionised IGM.

The \HeIII filling factor is modelled in a similar fashion; the much
higher energy photons ($>54.4\rm\,eV$) required to reionise \HeII mean
that this quantity can be decoupled\footnote{We have, however,
  ignored the effect of neutral helium on the evolution of the \HII
  filling factor, but the lower number density of helium, combined
  with the higher energy of the \HeI ionisation threshold, mean it
  will have only a small effect on \HI ionisation
  (e.g. Section~\ref{sec:ion_frac}).  For soft, stellar-like ionising
  spectra, \HII and \HeII ionisation fronts will furthermore closely
  trace each other during reionisation
  (\citealt{Friedrich12}). Lastly, note that \HeI ionisation is
  included in our radiative transfer simulations; the calculation here
  guides the choice of ionising emissivity in our simulations only. }
from \HI reionisation (e.g. \citealt{Madau99}).  Defining the
  comoving \HeII ionising emissivity as $\epsilon_{\rm HeII}$, we then
  have

\begin{equation} 
\frac{dQ_{\rm HeIII}}{dt} = \frac{ \epsilon_{\rm HeII}}{\langle n_{\rm He} \rangle} - Q_{\rm HeIII}C_{\rm HeIII}\frac{\langle n_{\rm e} \rangle_{\rm HeIII}}{a^{3}}\alpha_{\rm HeIII}(T), 
\label{eq:ffHe}
\end{equation}

\noindent
where $\langle n_{\rm He} \rangle = Y(1-Y)^{-1}\langle n_{\rm H}
\rangle/4$, $Y=0.258$ is the cosmic fraction of helium by mass,
$\langle n_{\rm e} \rangle_{\rm HeIII} = \langle n_{\rm H} \rangle + 2
\langle n_{\rm He} \rangle$, $C_{\rm HeIII} = \langle n_{\rm
  HeIII}^{2} \rangle/ \langle n_{\rm HeIII} \rangle^{2}$ and $\langle
n_{\rm e} \rangle_{\rm HII} = \langle n_{\rm H} \rangle + 2\langle
n_{\rm He} \rangle Q_{\rm HeIII}/Q_{\rm HII}$.  For a power-law
spectrum with spectral index $\alpha$, ${\epsilon}_{\rm HeII} =
4^{-\alpha}{\epsilon}_{\rm HI}$.  We shall assume $T=2\times
10^{4}\rm\,K$ and adopt time independent clumping factors $C_{\rm
  HII}=3$ and $C_{\rm HeIII}=3$ in Eqs~(\ref{eq:ffH}) and
(\ref{eq:ffHe}).  Note, however, the assumed clumping factor and
temperature are used as a guide only, and will be computed
self-consistently within our radiative transfer simulations.

\begin{figure}
\centering
\includegraphics[width=0.55\textwidth]{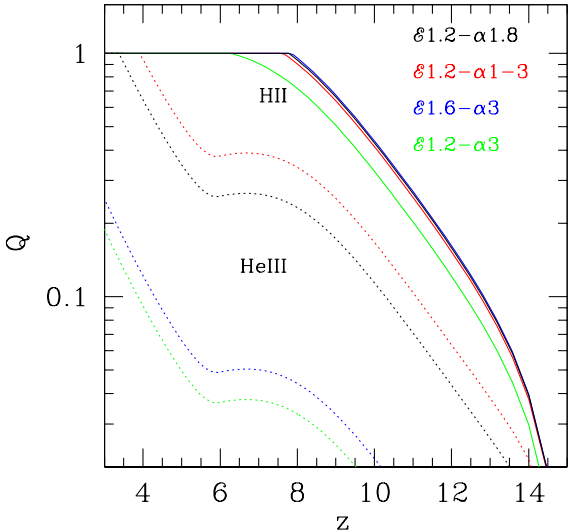}
\vspace{-1.3cm}
\caption{The evolution of the filling factor calculated for four of
  the reionisation models listed in Table 1. The black, red, blue and
  green curves correspond to model ${\mathcal E}$1.2-$\alpha$1.8,
  ${\mathcal E}$1.2-$\alpha$1-3, ${\mathcal E}$1.6-$\alpha$3 and
  ${\mathcal E}$1.2-$\alpha$3, while the solid and dotted curves
  display the \HII and \HeIII filling factors, respectively.}
\label{fig:ion_an}
\end{figure}

We next define the redshift evolution of the total comoving hydrogen
ionising emissivity as:

\begin{equation}
{\epsilon}_{\rm HI} = \left\{
\begin{array}{ll}
  {\mathcal E} \times 10^{50.89+{\rm log}(\chi(z))} \frac{\alpha^{-1}(\alpha_{\rm b}+3)}{2}              &    z>6,\\
  {\mathcal E} \times 10^{50.50-0.06(z-6)} \frac{\alpha^{-1}(\alpha_{\rm b}+3)}{2}           &    z \leq 6,\\
\end{array}\right.
\label{emiss}
\end{equation}
with
\begin{equation}
\chi(z) = \frac{\xi {\rm e}^{\zeta (z-9)}}{  \left( \xi - \zeta + \zeta {\rm e}^{\xi (z-9)} \right)}.
\label{emiss2}
\end{equation}

\noindent
Here ${\mathcal E}$ is a free parameter which sets the amplitude of
the emissivity, $\xi=14/15$, $\zeta=2/3$, $\alpha$ is the extreme-UV
(EUV) power-law spectral index of the sources and $\alpha_{\rm b}$ is
the spectral index of the ionising background; we shall assume the
same value for both. Eq.~(\ref{emiss}) is consistent with
observational constraints on the \HI photo-ionisation rate from the
\Lya forest at $z\leq 6$ (\citealt{BoltonHaehnelt07}, see also
Section~\ref{sec:sources}) and the mean free path\footnote{When the
  mean free path is much smaller than the horizon scale,
  $\epsilon_{\rm HI} \propto \Gamma_{\rm HI}\lambda_{\rm
    HI}^{-1}(\alpha_{\rm b}+3)\alpha^{-1}$, where $\Gamma_{\rm HI}$ is
  the \HI photo-ionisation rate and $\lambda_{\rm HI}$ is the mean
  free path of an ionising photon at the Lyman limit.  For a fixed
  photo-ionisation rate, a harder (softer) EUV spectral index or a
  smaller (larger) mean free path will therefore increase (decrease)
  the emissivity.}  for Lyman limit photons
(\citealt{SongailaCowie2010}), while Eq.~(\ref{emiss2})
(\citealt{SpringelHernquist03}) provides a simple parameterisation for
the rising emissivity at $z>6$ (peaking at $z=9$) required by the \Lya
forest data (e.g. \citealt{BoltonHaehnelt07,Pritchard09}).

We shall consider three different models for the spectral shape of the
ionising emission, all of which achieve \HI reionisation by $z \geq 6$
(i.e. $Q_{\rm HII}\sim 1$).  Our reference models (${\mathcal
  E}$1.2-$\alpha$1.8 and ${\mathcal E}$1.6-$\alpha$3) assume
$\alpha=1.8$ and $\alpha=3$, while a third model (${\mathcal
  E}$1.2-$\alpha$1-3) assumes 30 (70) per cent of the sources have
$\alpha=1$ (3).  A spectral index of $\alpha=1.8$ is typical of
quasars (\citealt{Telfer02}), while $\alpha=3$ is consistent with star
forming galaxies with metallicities close to solar, i.e. population-II
stellar sources (\citealt{Leitherer99}).  The third model assumes that a
fraction of the sources instead have rather hard spectra, $\alpha=1$,
typical of hard quasars or population-III stars
(e.g. \citealt{Bromm01}).

In Figure~\ref{fig:ion_an}, the evolution of both $Q_{\rm HII}$ (solid
curves) and $Q_{\rm HeIII}$ (dotted curves) is shown for model
${\mathcal E}$1.2-$\alpha$1.8 (black curves), ${\mathcal
  E}$1.2-$\alpha$1-3 (red curves) and ${\mathcal E}$1.6-$\alpha$3
(blue curves).  The models are normalised to have similar comoving
hydrogen ionising emissivities at each redshift, ensuring that any
differences in the reionisation histories are largely due to the
different EUV spectral indices.  For example, \HeII reionisation is
completed ($Q_{\rm HeIII}\sim 1$) progressively later in models
${\mathcal E}$1.2-$\alpha$1.8 and ${\mathcal E}$1.6-$\alpha$3, which
have softer ionising spectra compared to ${\mathcal
  E}$1.2-$\alpha$1-3.

Finally, in addition to these three reference reionisation histories,
we shall also consider two further models; ${\mathcal
  E}$1.2-$\alpha$1.8-H which excludes the treatment of helium, and
${\mathcal E}$1.2-$\alpha$3 which results in a late \HI reionisation
at $z \simeq 6$.  We include the latter to explore the possibility
that the volume weighted neutral fraction in the IGM at $z\simeq 7$
may be greater than 10 per cent. Such a substantial neutral
  fraction is suggested by recent observations, which, if confirmed by
  future investigations, may be in tension with models which satisfy
  constraints on the Thomson scattering optical depth and the hydrogen
  photo-ionisation rate (see Section~\ref{sec:observations} later for
further details).  The parameters for these reionisation models are
summarised in Table~\ref{table:models}.  Using these simple emissivity
models, we now turn to describing our cosmological radiative transfer
simulations.

\begin{table}
\caption{Summary of the ionising emissivity models used in this work.
  The columns indicate, from left to right, the name of the model, the
  amplitude of the emissivity, ${\mathcal E}$, the assumed EUV
  spectral index of the source spectrum, $\alpha$, and the percentage
  of sources with that spectrum, $f_\alpha$.  The final column
  indicates whether or not helium has been included in the
  simulations.}  \centering
\begin{tabular}{c c c c c} 
\hline
Model & ${\mathcal E}$ & $\alpha$ & $f_\alpha$ [\%] & He \\ 
\hline

${\mathcal E}$1.2-$\alpha$1.8-H   & 1.2  & 1.8   & 100     &  No  \\ 
${\mathcal E}$1.2-$\alpha$1.8     & 1.2  & 1.8   & 100     &  Yes \\ 
${\mathcal E}$1.2-$\alpha$1-3     & 1.2  & 1 (3) & 30 (70) &  Yes \\
${\mathcal E}$1.6-$\alpha$3       & 1.6  & 3     & 100     &  Yes     \\
${\mathcal E}$1.2-$\alpha$3       & 1.2  & 3     & 100     &  Yes    \\

\hline 
\end{tabular}
\label{table:models}
\end{table}


\section{Numerical simulations}
\subsection{Hydrodynamical simulations} 
\label{galform}

In order to perform our reionisation simulations, we require a model
for the intergalactic medium.  In this work we use a hydrodynamical
simulation performed in a comoving cubic box of size
$35.12h^{-1}\rm\,Mpc$.  The simulation was performed using the
parallel smoothed particle hydrodynamics (SPH) code {\small GADGET-3},
which is an updated version of the publicly available code {\small
  GADGET-2} (\citealt{Springel05}).  A total of $2\times 512^{3}$ dark
matter and gas particles were followed in the simulation, yielding a
mass per gas particle of $4.15\times 10^{6}h^{-1}\rm\,M_{\odot}$.
Beginning at $z=16$, outputs were obtained from the simulation at
redshift intervals $\Delta z = 0.5$ until $z=7$, and then at $\Delta z
= 0.4$ intervals until $z=5$.  Haloes were identified at each redshift
using a friend-of-friends halo finding algorithm with a linking length
of $0.2$.  Star formation was included using a simplified prescription
which converts all gas particles with overdensity $\Delta =
\rho/\langle \rho \rangle > 10^{3}$ and temperature $T<10^{5}\rm~K$
into collisionless stars.  Note that because of this simple treatment
our simulations do not self-consistently model star formation and
feedback. Instead, as discussed in Section~\ref{analytic}, we shall
model the ionising emissivity during reionisation using our
empirically motivated prescription.

The hydrodynamical simulation also includes the photo-ionisation and
heating of the IGM by a spatially uniform ionising background
(\citealt{HaardtMadau01}).  This model assumes the IGM is optically
thin, and that the IGM is reionised instantaneously at $z=9$.
Although we shall recompute the IGM ionisation and thermal state with
our radiative transfer (RT) simulations at all redshifts, including
the UV background in the hydrodynamical simulation at $z<9$ is
nevertheless important for properly modelling the gas distribution.
The photo-heating significantly reduces the clumping factor of the gas
in the hydrodynamical simulation due to pressure smoothing
(\citealt{Pawlik09}), and without this feedback effect the simulation
would over-predict the gas clumping factor towards the end of
reionisation.  On the other hand, we note that increasing the mass
resolution of our simulations would increase the clumping factor and
hence the rate of recombination in the simulations.  However, we defer
a detailed investigation of the clumping factor to a future study.
It should be noted though that, while the inclusion of a clumping factor
assures a better estimate of the gas recombination rate, it does not capture
all the relevant radiative transfer effects, such as self-shielding.

\subsection{Radiative transfer simulations}
\label{radtrans}

Once the hydrodynamical simulation outputs were obtained, the gas
number densities, $n$, temperatures, $T$ (at $z>9$ only, see
Section~\ref{galform}) and the halo masses, $M$, were transferred to a
$128^3$ grid for the RT calculations, which are performed as a
post-process.  The gridded densities and temperatures are obtained by
assigning the particle data to a regular grid using the SPH kernel
(e.g. \citealt{Monaghan92}).  The corresponding grid for the halo
masses is obtained by using the cloud-in-cell algorithm
(\citealt{Hockney88}) to assign the haloes identified by the
friends-of-friends algorithm to a regular grid with the same
dimensions.

The RT is followed using the code {\tt CRASH}
\citep{Ciardi_etal_2001,Maselli.Ferrara.Ciardi_2003,Maselli.Ciardi.Kanekar_2009,
  Partl_etal_2011}, which self-consistently calculates the evolution
of the hydrogen and helium ionisation state and the gas temperature.
{\tt CRASH} is a Monte Carlo based ray tracing scheme, where the
ionising radiation and its time varying distribution in space is
represented by multi-frequency photon packets which travel through the
simulation volume.  For further details regarding the radiative
transfer implementation we refer the reader to the original {\tt
  CRASH} papers.  For each output $i$ of the hydrodynamical
simulation, the RT is followed for a time $t_{\rm rt, i}=t_{\rm
  H}(z_{\rm i+1})-t_{\rm H}(z_{\rm i})$, where $t_{\rm H}(z_{\rm i})$
is the Hubble time corresponding to $z_{\rm i}$ which is the redshift
of output $i$.  The gas number density is updated at each
hydrodynamical simulation snapshot, and between two snapshots it is
evolved as $n(x_{\rm c},y_{\rm c},z_{\rm c})(z)=n(x_{\rm c},y_{\rm
  c},z_{\rm c})(z_{\rm i}) (1+z)^3/(1+z_{\rm i})^3$, where $(x_{\rm
  c},y_{\rm c},z_{\rm c})$ are the coordinates of cell $c$ and $z_{\rm
  i}>z>z_{\rm i+1}$.  Although the current implementation of {\tt
  CRASH} is able to model diffuse radiation without approximations, in
this work we choose to use the on-the-spot approximation.
The infinite velocity of light approximation is made and a  photon packet is considered as lost once it has exited the simulation
box, i.e. we do not use periodic boundary conditions.

The emission properties of the sources are derived as follows.  Guided
by our semi-analytical calculations in Section~\ref{galform}, we
assume that the total comoving hydrogen ionising emissivity at each
redshift is given by Eqs.~(\ref{emiss}) and (\ref{emiss2}).  Thus, the
total rate of ionising photons emitted at each output of the
hydrodynamical simulation is given by $\dot{N}_{\rm i}=\epsilon_{\rm
  HI}(z_{\rm i}) V_{\rm com}$, where $V_{\rm com}$ is the comoving
volume of the simulation.  The emissivity, $\dot{N}_{\rm i}$, is then
distributed among the sources according to their gas mass,
i.e. $\dot{N}_{\rm i,j}=\dot{N}_{\rm i} M_{\rm j}/M_{\rm tot,i}$,
where $j$ refers to the source and $M_{\rm tot,i}$ is the total gas
mass of sources at output $i$.  This method of assigning the
emissivity avoids assuming an escape fraction of ionising photons and
a star formation efficiency, which are very uncertain parameters.
Furthermore, as already discussed this empirical approach is designed
to be consistent with the existing observational constraints on the
photo-ionisation rate at $z\sim 6$.  Depending on the redshift
  and number of sources, we emit $10^5-10^6$ photon packets per source
  at each $t_{\rm rt,i}$, corresponding to a total of $\sim 5 \times
  10^7-10^{10}$ photon packets. At $z<8.5$ the total number is always
  $>10^9$, assuring convergence of the results to less than one
  percent (in relative terms) in the ionisation and neutral fraction
  for all the species, as well as the gas temperature (see the appendix
  for further details).

The ionisation fraction in the RT simulations is initialised to its
equilibrium value at $z_{\rm in}$, while the initial gas temperatures
correspond to those predicted by the hydrodynamical simulation, and
remain so until either a cell is crossed by a photon packet or at
redshifts $z<9$.  In the latter instance, the temperature is held
fixed at the $z=9$ value, prior to the onset of photo-heating in the
hydrodynamical simulation.  Once a cell is crossed by a photon packet,
the ionisation fraction and gas temperature are then updated
self-consistently within the radiative transfer calculation.

We have performed five RT simulations in total in this study, using
the models summarised in Table 1.  In order to assess the effect of
including helium on the evolution of hydrogen reionisation, in model
${\mathcal E}$1.2-$\alpha$1.8-H we include only hydrogen with a
fraction by mass (number) of 0.742 (0.92).  Furthermore, in model
${\mathcal E}$1.2-$\alpha$1-3, where there are two populations of
ionising sources with different power-law spectra, the EUV spectral
indices are assigned to sources randomly (i.e. no correlation with the
halo mass is assumed) to reproduce the correct relative proportions.
Note also that in all five models the power-law ionising spectra
extend to a maximum frequency of $\sim 200$~eV and that the
contribution from X-rays is not included.  Finally, due to the large
number of sources present in the box, to reduce the computational time
we adopt the clustering technique described and tested in Pierleoni et
al. (in prep).  This approach significantly speeds up our simulations;
for reference, the number of sources in the $35.12h^{-1}$ Mpc box is
reduced from 68 (80597) to 34 (14112) at $z=15$ (8).


\section{Empirical calibration of the reionisation simulations}\label{opticaldepth}

Before proceeding to discuss the results of our simulations in detail,
we first compare them to the two key observables we deliberately
calibrate to; the electron scattering optical depth and the background
photo-ionisation rate at $z \sim 6$ inferred from the \Lya forest.  As
mentioned in Section~\ref{analytic}, our choice for the reionisation
histories in the simulations is such that these key observational
constraints should automatically be satisfied.

\subsection{The Thomson scattering optical depth}

\begin{figure}
\centering \includegraphics[width=0.55\textwidth]{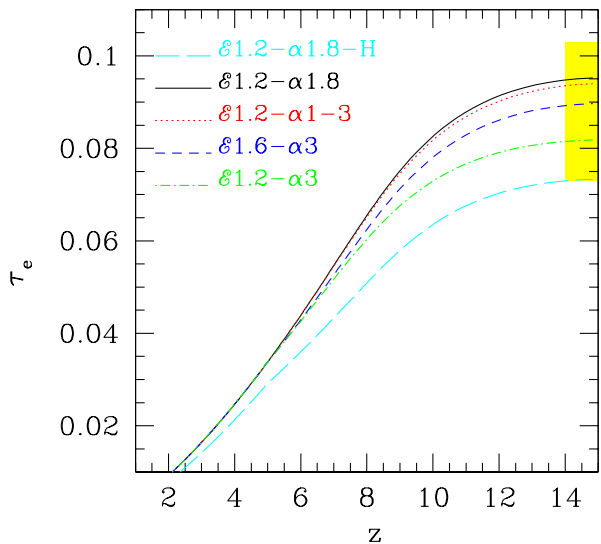}
\vspace{-1.3cm}
\caption{The Thomson scattering optical depth computed from each of
  our five radiative transfer simulations: ${\mathcal
    E}$1.2-$\alpha$1.8-H (cyan long dashed line), ${\mathcal
    E}$1.2-$\alpha$1.8 (black solid), ${\mathcal E}$1.2-$\alpha$1-3
  (red dotted), ${\mathcal E}$1.6-$\alpha$3 (blue dashed) and ${\mathcal
    E}$1.2-$\alpha$3 (green dotted dashed). The shaded area corresponds to
  the 7-yr {\tt WMAP} value of $0.088 \pm 0.015$ \citep{Komatsu11}.}
\label{fig:tau}
\end{figure}

We first consider the observational constraint on the integrated
reionisation history, in the form of the Thomson scattering optical
depth, $\tau_{\rm e}$.  In Figure~\ref{fig:tau} the evolution of
$\tau_e$ is shown for all five of our RT simulations, together with
the value measured by the 7-yr {\tt WMAP} mission, $0.088 \pm 0.015$
\citep{Komatsu11}.  The optical depth, $\tau_e$, is calculated from
the RT simulations as:

\begin{equation}
\tau_e= c \sigma_{\rm T} \int n_{\rm e}(t) dt,
\end{equation}

\noindent
where $c$ is the speed of light, $\sigma_{\rm T}=6.65 \times
10^{-25}$~cm$^2$ is the Thomson scattering cross section, $n_{\rm
  e}=n_{\rm HII}+n_{\rm HeII}+2n_{\rm HeIII}$ is the electron number
density in units of cm$^{-3}$ and $n_{\rm i}$ is the number density of
species $i$, with $i$=\HII, \HeII and He$\,\rm \scriptstyle III$. Here
$n_{\rm e}$ is evaluated directly from the simulations for $z>z_{\rm
  min}=5$, which is the redshift at which the radiative transfer
simulations are stopped.  At lower redshift, where we do not have
simulation outputs, we instead calculate $n_{\rm e}$ analytically
assuming that: {\it (i)} the average density equals the cosmological
mean density; {\it (ii)} hydrogen is completely ionised; {\it (iii)}
$x_{\rm {HeII}}=1$ ($x_{\rm {HeIII}}=0$) for $3<z<z_{\rm min}$ and
$x_{\rm {HeII}}=0$ ($x_{\rm {HeIII}}=1$) for $z<3$.

The Thomson scattering optical depth calculated in this manner has a
value of 0.073, 0.095, 0.094, 0.090, 0.081 for the simulations
${\mathcal E}$1.2-$\alpha$1.8-H, ${\mathcal E}$1.2-$\alpha$1.8,
${\mathcal E}$1.2-$\alpha$1-3, ${\mathcal E}$1.6-$\alpha$3 and
${\mathcal E}$1.2-$\alpha$3, respectively.  As expected, these values
are consistent with those measured by the {\tt WMAP} satellite
\citep{Komatsu11}.  Note, however, that for model ${\mathcal
  E}$1.2-$\alpha$1.8-H we consider only the contribution from
hydrogen.  The inclusion of helium in these models is clearly
important, adding an additional $\tau_{\rm e}\sim 0.02$ to the total
optical depth for ${\mathcal E}$1.2-$\alpha$1.8.  This is largely
because of the extra electrons liberated by the reionisation of
helium, but will also be partly due to the higher IGM temperatures
which arise from \HeII photo-heating; the temperature dependence of
the \HII recombination rate, $\alpha_{\rm HII}\propto T^{-0.7}$, means
higher temperatures will produce a slight increase in the \HII
fraction and hence the electron number density.

\subsection{The background photo-ionisation rate}

\begin{figure}
\centering \includegraphics[width=0.47\textwidth]{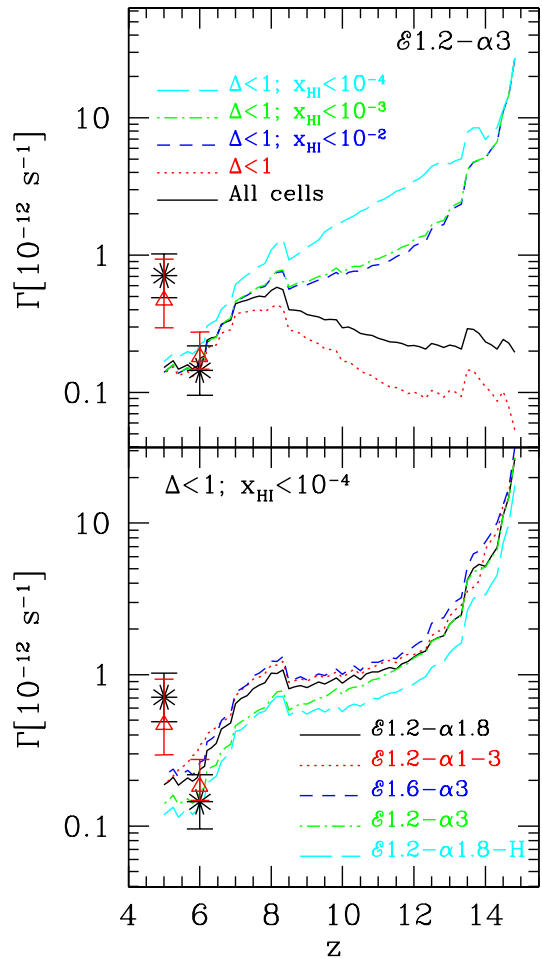}
\caption{ {\it Upper panel:} The redshift evolution of the volume
  averaged photo-ionisation rate, $\Gamma_{\rm HI}$, for model
  ${\mathcal E}$1.2-$\alpha$3.  The black solid curve shows the
  photo-ionisation rate for all cells, while the dotted red curve
  displays the data for underdense cells ($\Delta<1$) only.  The
  remaining three curves show the photo-ionisation rate in underdense
  cells, but now with the additional condition that $x_{\rm HI}<10^{-2}$
  (blue dashed), $10^{-3}$ (green dot-dashed) and $10^{-4}$ (cyan long
  dashed).  {\it Lower panel:} The redshift evolution of the volume
  averaged $\Gamma_{\rm HI}$ for cells with overdensity $\Delta<1$ and
  $x_{\rm HI}<10^{-4}$ only. The curves correspond to the models
  ${\mathcal E}$1.2-$\alpha$1.8-H (cyan long dashed lines), ${\mathcal
    E}$1.2-$\alpha$1.8 (black solid), ${\mathcal E}$1.2-$\alpha$1-3
  (red dotted), ${\mathcal E}$1.6-$\alpha$3 (blue dashed) and
  ${\mathcal E}$1.2-$\alpha$3 (green dotted-dashed).  In all panels
  the triangles and stars display respectively the observational
  constraints from the \Lya forest \citep{Wyithe.Bolton_2011} and the
  proximity effect \citep{Calverley11}.
}
\label{fig:gamma}
\end{figure}

The photo-ionisation rates are compared to the observational data in
Figure~\ref{fig:gamma}.  This comparison, however, is less
straightforward for two reasons.  Firstly, the photo-ionisation rate
is not a direct output from our RT simulations, and so we must
estimate it indirectly by assuming ionisation equilibrium in each cell
$(x_{\rm c},y_{\rm c},z_{\rm c})$, such that:

\begin{equation}
\Gamma_{\rm HI}=\alpha_{\rm HII}(T) \frac {n_{\rm e} n_{\rm HII}} {n_{\rm HI}} - \gamma_{\rm eHI}(T) n_{\rm e}, \label{eq:PIrate}
\end{equation}

\noindent
where $\alpha_{\rm HII}$ and $\gamma_{\rm eHI}$ are the hydrogen
recombination and collisional ionisation rate in units of $\rm
cm^3\,s^{-1}$, respectively. All the other quantities have their usual
meaning.  This will be a reasonable approximation for most of the
cells in our simulation volume after they have been reionised, but
will break down close to reionisation when non-equilibrium effects are
important.  Secondly, the observational constraints on the
photo-ionisation rate are derived from the \Lya absorption observed in
$z\simeq 6$ quasar spectra
(e.g. \citealt{Fan06,BoltonHaehnelt07,Calverley11}).  The transmitted
\Lya flux at these redshifts preferentially samples highly ionised,
underdense regions in the IGM, and so we must take care to use similar
criteria when comparing to volume averaged values in the simulations.

In the upper panel of Figure~\ref{fig:gamma} the evolution of the
volume averaged \HI photo-ionisation rate, $\Gamma_{\rm HI}$, is shown
for model ${\mathcal E}$1.2-$\alpha$3. The different curves display
$\Gamma_{\rm HI}$ for a variety of different sub-samples drawn from
the simulation volume.  The black solid curve shows the
photo-ionisation rate for all cells, whereas the dotted red curve
displays the data for underdense cells ($\Delta<1$) only.  The
remaining three curves again show the photo-ionisation rate in
underdense cells, but now with the additional condition that $x_{\rm
  HI}<10^{-2}$ (blue dashed), $10^{-3}$ (green dot-dashed) and
$10^{-4}$ (cyan long dashed).  These cuts correspond to $\sim 0.13$,
$0.13$, $0.84$ per cent of the total number of cells in the simulation
volume at $z=14$. At $z=6$ the percentages are instead 63, 62, and 18,
respectively.  When all cells are included, the evolution of
$\Gamma_{\rm HI}$ rises to a peak at $z \sim 8$ (following the rising
emissivity at $z>6$ in Eq.~\ref{emiss2}) but declines toward higher
redshift.  This is because a larger number of neutral cells are
present toward higher redshifts, lowering the volume averaged
photo-ionisation rate. The average photo-ionisation rate is slightly
lower if only underdense cells are included because the overdense (and
hence first to reionise) regions are discarded.  In other words, the
photo-ionisation rates are higher in the overdense cells since the
ionising radiation is correlated with the underlying density field
(see also \citealt{Iliev08,MesingerFurlanetto09}).

At $z=6$, by which time all the underdense regions in the simulation
have been reionised, all curves converge to a similar value. Note,
however, that in the cases where cuts in the neutral fraction are also
applied, at $z>6$ the photo-ionisation rate is always higher compared
to the average for all the underdense cells (red dotted curve).  This
is in part because the averages are, by definition, only for highly
ionised cells which are assumed to be in ionisation equilibrium.  The
difference is more pronounced at $z>8$, however, when the ionised
regions probed are the increasingly rare ionised bubbles around
sources.  We thus also expect higher photo-ionisation rates because
the selected cells are closer to the ionising sources.  However, these
regions are rare and so only provide a small contribution to the
overall volume averaged ionisation rate.

In the lower panel of Figure~\ref{fig:gamma} the evolution of the
volume averaged $\Gamma_{\rm HI}$ is shown for all five simulations in
underdense cells which are highly ionised only ($x_{\rm HI}<10^{-4}$).
Note that this cut most closely represents the regions of the IGM from
which the photo-ionisation rates are measured at $z\simeq 6$
(\citealt{BoltonHaehnelt07}).  The redshift evolution of $\Gamma_{\rm
  HI}$ is, as might be expected, similar for all models.  Model
${\mathcal E}$1.2-$\alpha$3 typically gives a smaller photo-ionisation
rate due to the lower normalisation of the emissivity.  On the other
hand, model ${\mathcal E}$1.2-$\alpha$1.8-H always has a slightly
lower value of $\Gamma_{\rm HI}$ compared to the case including
helium, ${\mathcal E}$1.2-$\alpha$1.8.  Note, however, the
photoionisation rates are inferred from Eq.~(\ref{eq:PIrate}) rather
than directly obtained, and so variations in the gas temperature and
electron number density in this model will be partly responsible for
this difference.

Finally, as required, we find that for all models at $z=6$ the
photo-ionisation rates are consistent with the observational
constraints from the \Lya forest (\citealt{Wyithe.Bolton_2011}) and
proximity effect (\citealt{Calverley11}), represented by triangles and
stars with error bars in Figure~\ref{fig:gamma}, respectively.  On the
other hand, the photo-ionisation rates at $z=5$ underpredict the
observed values by a factor of $2$--$3$, despite the fact we have
deliberately used an ionising emissivity which agrees with these data
when assuming a mean free path consistent with recent observational
measurements (e.g. \citealt{SongailaCowie2010}).  This discrepancy may
be understood by recalling that $\Gamma_{\rm HI} \propto \epsilon_{\rm
  HI}\lambda_{\rm HI}$, where $\lambda_{\rm HI}$ is the mean free path
at the Lyman limit.  Assuming a power-law slope for the \HI column
density distribution of $\beta=1.3$, \cite{SongailaCowie2010} measure
$\lambda_{\rm HI}\simeq 84$ (49) comoving Mpc at $z=5$ (6).  In
comparison, our simulation volume is $48.7$ comoving Mpc on a side.
This sets an effective upper limit on the mean free path of ionising
photons in our simulations which is around half the observed value at
$z=5$.  Our small simulation box therefore most likely accounts for
this apparent discrepancy, and we caution that the ionising emissivity
in our simulations is underestimated at $z<6$ as a result.


\section{The evolution of the IGM ionisation and thermal state}

We have found that our simulations are in reasonable agreement with
both the observed Thomson scattering optical depth and background
photo-ionisation rate at $z=6$, giving us confidence that we may now
explore the implications of these models for the ionisation and
thermal state of the IGM in further detail.

\begin{figure}
\centering \includegraphics[width=0.45\textwidth]{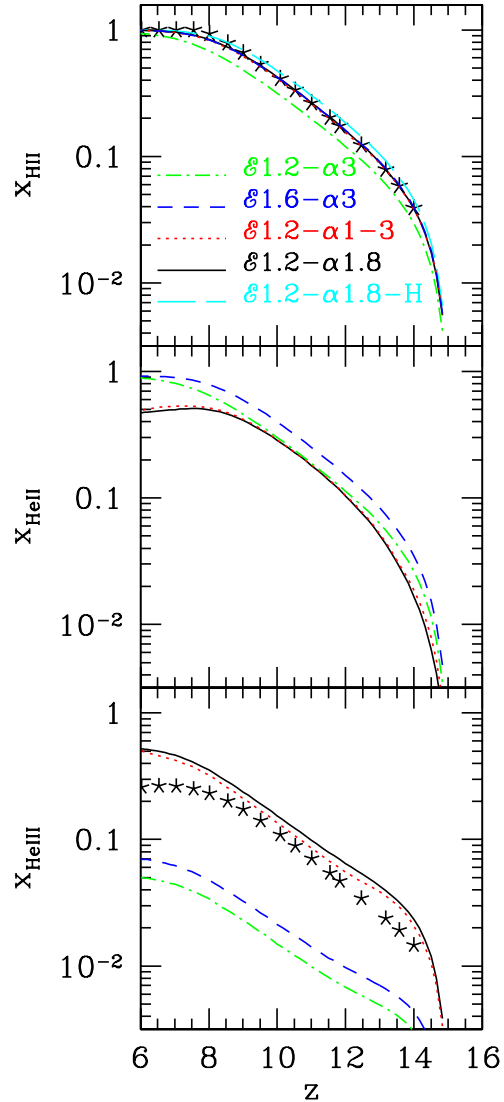}
\caption{ {\it Upper panel:} The evolution of the volume averaged \HII
  fraction calculated with the radiative transfer simulations for
  models ${\mathcal E}$1.2-$\alpha$1.8-H (long dashed cyan line),
  ${\mathcal E}$1.2-$\alpha$1.8 (solid black lines), ${\mathcal
    E}$1.2-$\alpha$1-3 (dotted red), ${\mathcal E}$1.6-$\alpha$3
  (dashed blue) and ${\mathcal E}$1.2-$\alpha$3 (dot-dashed green).
  Note the solid black, dotted red, and dashed blue lines are almost
  indistinguishable. The stars display the semi-analytic result for
  model ${\mathcal E}$1.2-$\alpha$1.8 (see Section 2.2 for details).
  {\it Middle panel:} As for the upper panel but for the volume
  averaged \HeII fraction.  Note that in this case model ${\mathcal
    E}$1.2-$\alpha$1.8-H is not present.  {\it Lower panel:} As for
  the upper middle but for the volume averaged \HeIII fraction. The
  stars again refer to the semi-analytic result for model ${\mathcal
    E}$1.2-$\alpha$1.8. }
\label{fig:ion_sim}
\end{figure}

\subsection{The ionisation fraction}
\label{sec:ion_frac}

The volume averaged ionisation fractions predicted by the RT
simulations are displayed in Figure~\ref{fig:ion_sim}, where the
upper, middle and lower panels refer, respectively, to the evolution
of the H$\,\rm \scriptstyle II$, \HeII and \HeIII fractions for the models
summarized in Table~\ref{table:models}. 
Reionisation is largely complete by $z=7$
in all models (i.e. $x_{\rm HI} \leq 0.05$), with the exception of
${\mathcal E}$1.2-$\alpha$3, which has an \HI fraction of 0.15 at
$z=7$.

Although the aim of this study is not to compare the RT simulations
with the semi-analytic calculations used to guide our choice of
ionising emissivity, it is interesting to note that the numerical
models reproduce the semi-analytic results for the \HII evolution
remarkably well.  However, the agreement is to some extent a fortunate
coincidence; a different assumption for the hydrogen clumping factor
in \HII regions, $C_{\rm HII}$, or IGM temperature in the
semi-analytical model would worsen the agreement.  The agreement
between the numerical and semi-analytical evolution of $x_{\rm HeIII}$
is slightly poorer, which is indeed most likely due to slightly
different values for the clumping factor and/or temperature in the two
approaches.  Nevertheless, the general agreement indicates that
semi-analytical approaches are indeed useful for quickly exploring
parameter space in reionisation models, at least in terms of the
volume of the IGM which is ionised.  This is perhaps not too
surprising; both calculations are effectively just counting ionising
photons and recombinations.  Indeed, ``semi-numerical'' schemes which
additionally follow the topology of reionisation are also in
relatively good agreement with the results of full RT calculations
(e.g. \citealt{Zahn11}).

The long dashed cyan curve in the top panel of
Figure~\ref{fig:ion_sim} compares the ${\mathcal E}$1.2-$\alpha$1.8-H
model, which excludes helium, to the corresponding reference run
${\mathcal E}$1.2-$\alpha$1.8.  The abundance of
\HII in ${\mathcal E}$1.2-$\alpha$1.8-H is slightly higher because all
of the ionising photons ($>13.6\,\rm eV$) are used to ionise hydrogen.
The inclusion of helium in model ${\mathcal E}$1.2-$\alpha$1.8 has a
small effect on the evolution of the neutral hydrogen fraction, as
some of the hydrogen ionising photons with energies $>24.6\,\rm eV$
are now used to reionise He$\,\rm \scriptstyle I$.  However, the
difference between $x_{\rm HII}$ in the ${\mathcal
  E}$1.2-$\alpha$1.8-H and ${\mathcal E}$1.2-$\alpha$1.8 models is
never above a few per cent.

\begin{figure*}
\includegraphics[width=1.00\textwidth]{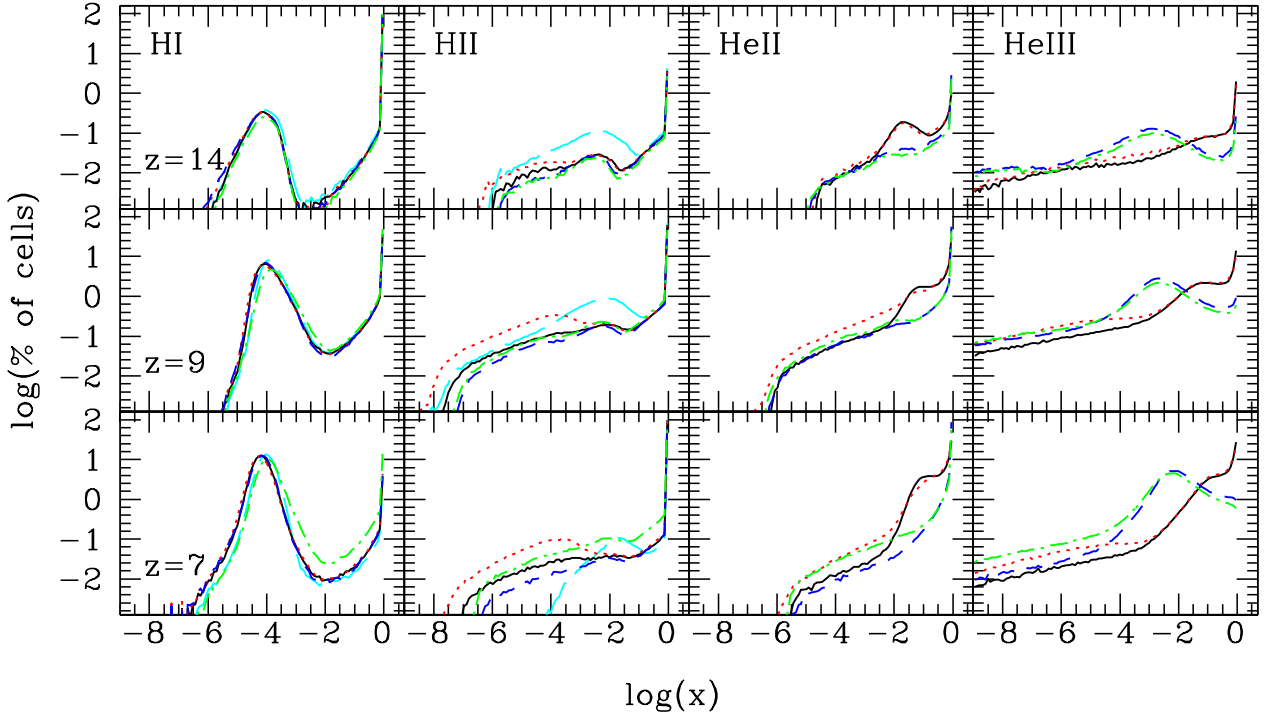}
\vspace{-1cm}
\caption{The percentage of cells in the radiative transfer simulations
  as a function H$\,\rm \scriptstyle I$, H$\,\rm \scriptstyle II$,
  He$\,\rm \scriptstyle II$ and He$\,\rm \scriptstyle III$ fractions
  (from left to right) at $z$=14 (upper row), 9 (middle row) and 7
  (lower row).  The curves in each panel correspond to a different
  reionisation history: ${\mathcal E}$1.2-$\alpha$1.8-H (long dashed
  cyan, first two columns only), ${\mathcal E}$1.2-$\alpha$1.8 (solid
  black), ${\mathcal E}$1.2-$\alpha$1-3 (dotted red), ${\mathcal
    E}$1.6-$\alpha$3 (dashed blue) and ${\mathcal E}$1.2-$\alpha$3
  (dot-dashed green).}
\label{fig:distr_x}\end{figure*}

The impact of different spectral energy distributions on the ionised
fractions can be seen by comparing model ${\mathcal E}$1.2-$\alpha$1.8
to models ${\mathcal E}$1.2-$\alpha$1-3 and
${\mathcal E}$1.6-$\alpha$3.  Interestingly, for
the mixed source model ${\mathcal E}$1.2-$\alpha$1-3, all three
ionisation fractions (H$\,\rm \scriptstyle II$, He$\,\rm \scriptstyle
II$, He$\,\rm \scriptstyle III$) are extremely similar to those of
model ${\mathcal E}$1.2-$\alpha$1.8.  This is because both the
comoving emissivity and the number of photons with frequencies above
the helium ionisation thresholds are very similar in the two models.
Spectra with power-law indices $\alpha=1$, 1.8 and 3 have a percentage
of ionising photons above the \HeI (He$\,\rm \scriptstyle II$)
ionisation threshold, i.e. above 24.6~eV (54.4~eV), of $\sim 52$
(19.5), 34 (7.5) and 17 (1.5) per cent, respectively.  In the case of
the models with the softer ionising spectrum (i.e. $\alpha=3$),
$x_{\rm HeIII}$ is much lower due to the paucity of higher energy
photons.  The softer spectrum is also reflected in the evolution of
$x_{\rm HeII}$, which is very similar to that of $x_{\rm HII}$.  

Finally, model ${\mathcal E}$1.2-$\alpha$3 exhibits very similar
behaviour to that of ${\mathcal E}$1.6-$\alpha$3 because they have the
same spectral index, but the ionisation fractions at the same redshift
are smaller due to the lower amplitude of the comoving emissivity.
Note, however, that both of these models have EUV spectral indices
which are too soft to complete \HeII reionisation by $z\simeq
2.5$--$3$ (e.g. Fig.~\ref{fig:ion_an}).  These models are therefore
likely inconsistent with the \HeII \Lya forest data at $z\simeq 3$
(e.g. \citealt{Shull10,Worseck11,Syphers11}) unless the ionising
background spectral shape hardens at $z<6$, perhaps due to the
increasing contribution of quasars to the ionising background.  For
reference, the volume averaged ionisation fractions at $z=14,\, 9,\,
7$ and 6 are summarised in Table~\ref{table:xt}.

\begin{table}
\caption{Summary of the volume averaged ionisation fractions and
  temperature within the RT simulations.  The columns indicate, from
  left to right, the name of the model, the redshift $z$, the volume
  averaged ionisation fractions $x_{\rm HII}$, $x_{\rm HeII}$ and
  $x_{\rm HeIII}$, and the volume averarged temperature $T$.}
\centering
\begin{tabular}{cccccc}        
\hline
Model & $z$ & $x_{\rm HII}$ & $x_{\rm HeII}$ & $x_{\rm HeIII}$ & $T$ [K] \\
\hline
      & 14  & 0.045        & --             &   --            & 918     \\
      & 9   & 0.695        & --             &   --            & 9760    \\[-1ex]
\raisebox{1.5ex}{${\mathcal E}$1.2-$\alpha$1.8-H}
      & 7   & 0.981        & --             &   --            & 11047   \\
      & 6   & 0.998        & --             &   --            & 10224   \\[1ex]
\hline
      & 14  & 0.038        & 0.017          &   0.023         & 820     \\      
      & 9   & 0.632        & 0.410          &   0.238         & 10464   \\[-1ex]
\raisebox{1.5ex}{${\mathcal E}$1.2-$\alpha$1.8}
      & 7   & 0.960        & 0.499          &   0.464         & 16594   \\      
      & 6   & 0.993        & 0.472          &   0.522         & 16998   \\[1ex]
\hline
      & 14  & 0.038        & 0.019          &   0.021         & 804     \\            
      & 9   & 0.618        & 0.419          &   0.211         & 10190   \\[-1ex]
\raisebox{1.5ex}{${\mathcal E}$1.2-$\alpha$1-3}  
      & 7   & 0.953        & 0.531          &   0.425         & 16565   \\            
      & 6   & 0.993        & 0.490          &   0.504         & 17454   \\[1ex]
\hline
      & 14  & 0.039        & 0.035          &   0.004         & 643     \\                  
      & 9   & 0.627        & 0.594          &   0.032         & 7674    \\[-1ex]
\raisebox{1.5ex}{${\mathcal E}$1.6-$\alpha$3}  
      & 7   & 0.957        & 0.894          &   0.063         & 11425   \\                  
      & 6   & 0.993        & 0.922          &   0.070         & 11347   \\[1ex]
\hline
      & 14  & 0.029        & 0.026          &   0.003         & 488     \\              
      & 9   & 0.481        & 0.459          &   0.023         & 6020    \\[-1ex]
\raisebox{1.5ex}{${\mathcal E}$1.2-$\alpha$3}  
      & 7   & 0.852        & 0.807          &   0.044         & 10643   \\         
      & 6   & 0.938        & 0.888          &   0.050         & 11347   \\[1ex]
\hline
\end{tabular}
\label{table:xt}
\end{table}

A more quantitative representation of the distributions of the various
ionised fractions is displayed in Figure~\ref{fig:distr_x}, where from
left to right the percentage of cells as a function of $x_{\rm HI}$,
$x_{\rm HII}$, $x_{\rm HeII}$ and $x_{\rm HeIII}$ are shown for the
five reionisation models at $z=14$ (upper row), 9 (middle row) and 7
(lower row).  At the highest redshifts most of the hydrogen is in a
neutral state, but as the redshift decreases and reionisation proceeds
the percentage of ionised cells increases for all models. During the
final stages of reionisation (represented here at $z=7$), most of the
cells will be fully or almost fully ($x_{\rm HII}>0.9$) ionised and,
as a consequence, the percentage of cells with a lower ionisation
fraction decreases again.  Model ${\mathcal E}$1.2-$\alpha$1.8-H
generally has a slightly
higher number of highly ionised cells compared to the three reference
models.  This is again because helium is absent in this model; all the
ionising photons are thus absorbed by hydrogen, enabling hydrogen
reionisation to proceed slightly more quickly. The behaviour of
models ${\mathcal E}$1.6-$\alpha$3 and ${\mathcal
  E}$1.2-$\alpha$1-3 is also rather similar to ${\mathcal
  E}$1.2-$\alpha$1.8, except ${\mathcal E}$1.6-$\alpha$3
(${\mathcal E}$1.2-$\alpha$1-3) has slightly less (more) cells with
very small ionised fractions.  This is because of the softer (harder)
ionising spectra which produce proportionally more (less) hydrogen
ionising photons.  As noted previously, the \HeII and \HeIII
ionisation fractions for ${\mathcal E}$1.2-$\alpha$1-3
and ${\mathcal E}$1.2-$\alpha$1.8 show rather similar
behaviour, while ${\mathcal E}$1.6-$\alpha$3 exhibits
much smaller \HeIII fractions due to the presence of fewer hard,
helium ionising photons.  A situation similar to model ${\mathcal
  E}$1.6-$\alpha$3 applies to ${\mathcal E}$1.2-$\alpha$3,
with the difference that the lower emissivity means
reionisation is less advanced.

From this analysis it is clear that including intergalactic helium and
a treatment of multi-frequency radiative transfer has a rather small
effect on the ionisation state of hydrogen during reionisation.
However, the hard ionising photons capable of ionising helium will
also significantly photo-heat the IGM. We therefore now turn to
consider the effect on the thermal state of the IGM at high redshift.

\subsection{The volume averaged temperature}

\begin{figure}
  \begin{center}
    \includegraphics[width=0.45\textwidth]{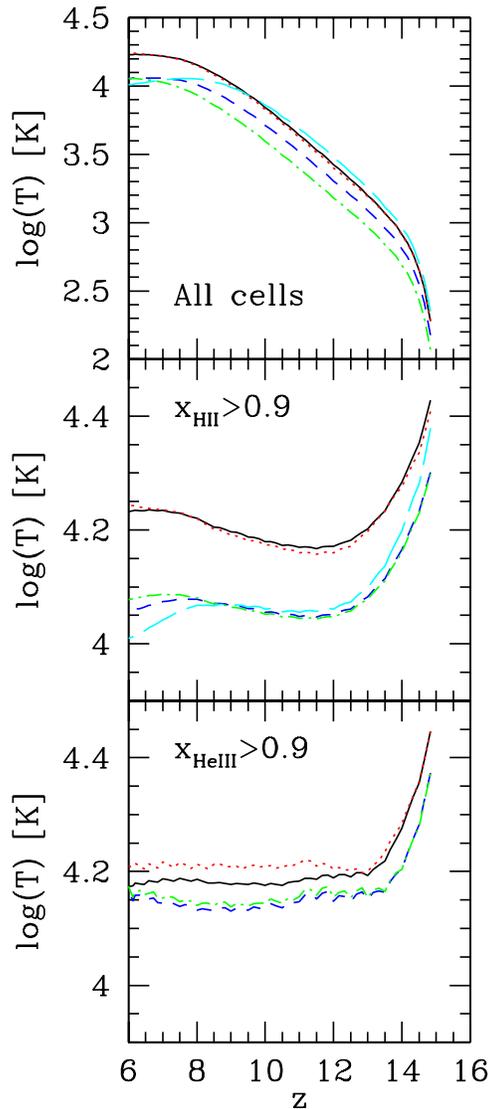}
    \caption{Redshift evolution of the volume averaged temperature.
      The curves correspond to models ${\mathcal E}$1.2-$\alpha$1.8-H
      (long dashed cyan), ${\mathcal E}$1.2-$\alpha$1.8 (solid black),
      ${\mathcal E}$1.2-$\alpha$1-3 (dotted red), ${\mathcal
        E}$1.6-$\alpha$3 (dashed blue) and ${\mathcal E}$1.2-$\alpha$3
      (dotted-dashed green) respectively.  {\it Upper panel}: The
      temperature calculated by averaging over all
      the cells in the simulation volume.  {\it Middle panel}: The
      temperature calculated by averaging over only those cells with
      $x_{\rm HII}>0.9$.  {\it Lower panel}: The temperature
      calculated by averaging over only on those cells with $x_{\rm
        HeIII}>0.9$.}
    \label{fig:temp}
  \end{center}
\end{figure}

The redshift evolution of the volume averaged gas temperature in our
five reionisation models is displayed in the upper panel of
Figure~\ref{fig:temp}.  This quantity will depend on the volume of the
IGM already reionised at any given redshift, as well as the spectral
shape of the sources in the simulation and whether or not helium
photo-heating is included.  The first point to note is that at early
times ($z>10$) model ${\mathcal E}$1.2-$\alpha$1.8-H 
has a volume averaged gas temperature which is $\sim$10 per
cent higher than the corresponding model with helium, ${\mathcal
  E}$1.2-$\alpha$1.8.  This is due to the slightly
larger volume of the IGM in which hydrogen is photo-ionised and heated
compared to the other models.  This arises from the fact (as discussed
earlier) that no hydrogen ionising photons are used to ionise neutral
helium.  Note, however, that by $z \sim 10$ the inclusion of \HeII
photo-ionisation results in a higher average temperature for
${\mathcal E}$1.2-$\alpha$1.8 compared to ${\mathcal
  E}$1.2-$\alpha$1.8-H.  In addition, in the absence of any additional
heating from \HeII photo-ionisation, the temperature for ${\mathcal
  E}$1.2-$\alpha$1.8-H slightly declines at $z<9$ as the IGM cools.

The volume averaged temperature evolution does not exhibit any
substantial difference between models ${\mathcal E}$1.2-$\alpha$1.8
and ${\mathcal E}$1.2-$\alpha$1-3, which is
expected from the very similar behaviour of the ionisation fractions
discussed earlier. On the other hand, despite having a similar
behaviour for the evolution of the \HII filling factor, the softer
ionising spectrum used by ${\mathcal E}$1.6-$\alpha$3 
produces temperatures 20-30 per cent lower than ${\mathcal
  E}$1.2-$\alpha$1.8.  This is partly because the volume filling
factor of \HeIII is smaller in this model, but also because the softer
spectrum results in less energy (and hence photo-heating) per
photo-ionisation on average.  Lastly, for the case of ${\mathcal
  E}$1.2-$\alpha$3, the volume averaged
temperature is $\sim$20--25 per cent lower compared to model ${\mathcal
  E}$1.6-$\alpha$3 over most of reionisation, but converges to a
similar temperature by $z=6$.  This is due to the lower ionising
emissivity, and hence smaller filling factor of ionised hydrogen, used
in model ${\mathcal E}$1.2-$\alpha$3 which delays the completion of
hydrogen reionisation to $z \simeq 6$.

\begin{figure*}
  \centering
\includegraphics[width=0.95\textwidth]{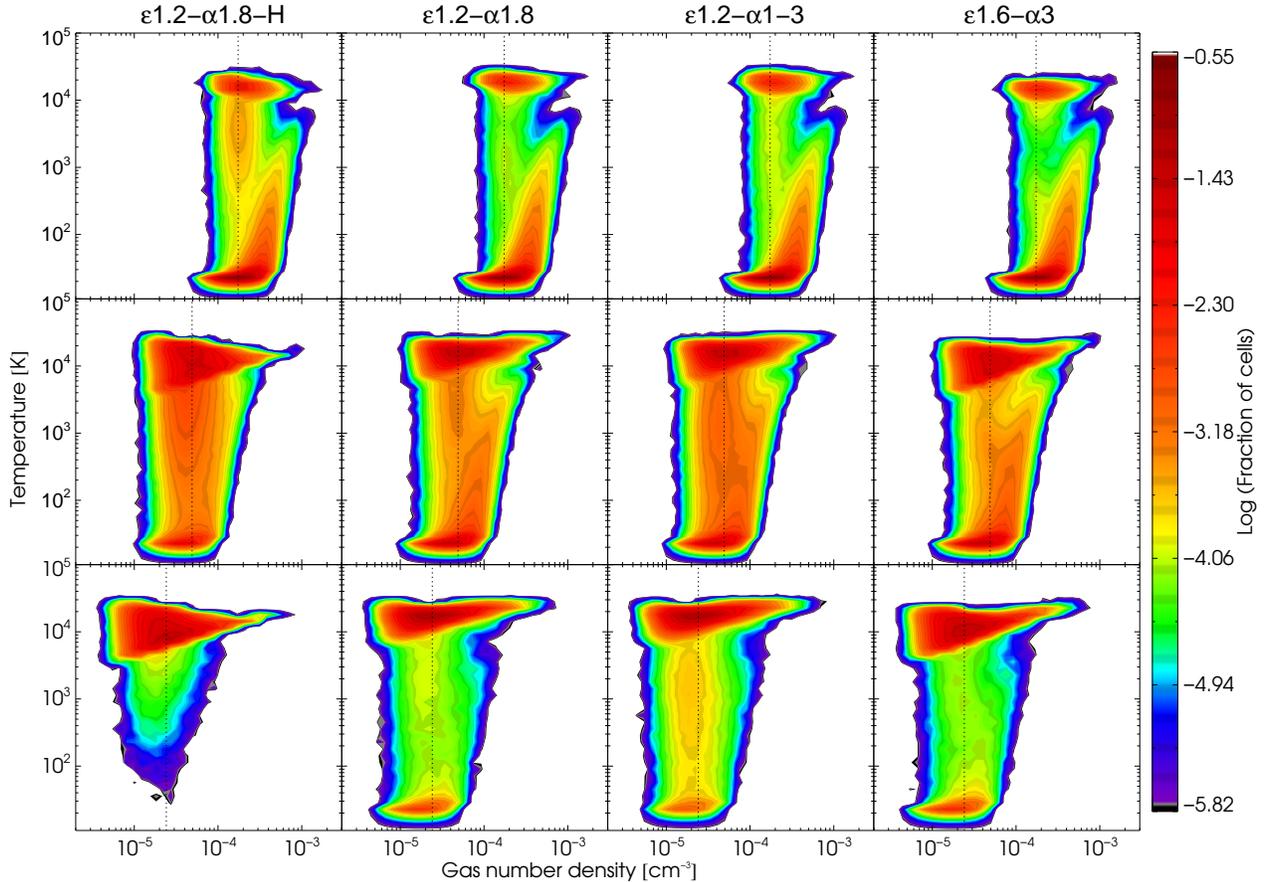}
\vspace{-0.4cm}
\caption{Contour plots of the distribution of gas temperature against
  proper number density for models ${\mathcal E}$1.2-$\alpha$1.8-H,
  ${\mathcal E}$1.2-$\alpha$1.8, ${\mathcal E}$1.2-$\alpha$1-3 and
  ${\mathcal E}$1.6-$\alpha$3 (from left to right). The colour scale
  corresponds to the percentage of cells within each contour.  The rows
  refer to redshift $z$=14, 9 and 7 (from top to bottom).  The dashed
  vertical lines correspond to the average density in the box.}
\label{fig:contours_temp}
\end{figure*}

We can also isolate the effect of the source spectrum from the volume
filling factor of ionised regions by calculating the volume averaged
temperature in \HII and \HeIII regions only, i.e. in regions with
$x_{\rm HII}>x_{\rm min}$ (middle panel) and $x_{\rm HeIII}>x_{\rm
  min}$ (lower panel), where $x_{\min}=0.9$.  We have verified that
varying our choice of threshold results in similar average
temperatures as long as $x_{\min}>0.1$.  The gas temperature reaches
its maximum value in the \HII and \HeIII regions at the highest
redshift, when only a small percentage of cells ($<1$ per cent) in the
vicinity of the first sources have been reached by ionising photons
and there has been very little time for the gas to cool.  As
reionisation proceeds, more cells are ionised, but those that have
been ionised earlier start to cool primarily by adiabatic expansion
(for gas close to mean density) and Compton scattering. The net result
is the average temperature in \HII regions decreases until $z \sim 12$
(when $\sim 5$ per cent of the cells have $x_{\rm HII}>x_{\rm min}$).
At lower redshifts, an increase in the number of cells in \HII regions
which have also experienced \HeII photo-heating, combined with the
fact that more cells are being reionised per unit time with the
increasing emissivity, results in the volume averaged \HII region
temperatures gradually increasing again toward $z=6$.  Note, however,
that for model ${\mathcal E}$1.2-$\alpha$1.8-H, where \HeII heating is
absent, the temperature starts to fall again at $z<8$ once \HI
reionisation is complete and the ionising emissivity begins to decline. 

The behaviour of the volume averaged temperature in the \HeIII regions
(lower panel) is broadly similar to the case for \HII regions, with a
high initial temperature followed by cooling.  However, in this
instance the temperature remains almost constant at $z<12$.  Here the
effect of cooling is offset by the temperature increase due to freshly
ionised \HeIII regions which continue to grow at $z<6$.  Finally, note
that for both the \HII and \HeIII regions, models ${\mathcal
  E}$1.2-$\alpha$1.8 and ${\mathcal E}$1.2-$\alpha$1-3 always exhibit
higher temperatures compared to the other models because of the energy
input from hard photons during \HeII photo-heating.  This is of
  particular relevance when comparisons with observations are made,
  and will be further discussed in Section~\ref{sec:observations}.

\subsection{The IGM temperature-density relation}

The temperatures in the simulations are examined in more detail in
Figure~\ref{fig:contours_temp}, which displays the distribution of the
gas temperature versus the proper number density for ${\mathcal
  E}$1.2-$\alpha$1.8-H, ${\mathcal E}$1.2-$\alpha$1.8, ${\mathcal
  E}$1.2-$\alpha$1-3 and ${\mathcal E}$1.6-$\alpha$3 (from left to
right). From top to bottom, each row displays the temperature-density
plane at redshift $z$=14, 9 and 7.  For reference, the volume averaged
temperatures at $z$=14, 9 and 7 for all models are given in
Table~\ref{table:xt}.  All cases show common features.  While
initially most of the neutral gas lies along a cold ($\sim$ 25~K)
isothermal locus, as reionisation proceeds more cells are photo-heated
into a second, multi-valued grouping at higher temperature.  At
$z=14$, a plume of hotter gas extending out to $T\simeq 10^{3}\rm\,K$
from the cold grouping toward higher densities is clearly apparent;
this is due to shocked heated gas in the hydrodynamical
simulation. Towards the end of reionisation, the vast majority of
cells have reached their maximum temperature, which depends primarily
on the ionising spectrum adopted.  The fact that ionisation proceeds
at a faster pace in model ${\mathcal E}$1.2-$\alpha$1.8-H is reflected
by the temperature behaviour: while at $z=7$ almost all the cells in
case ${\mathcal E}$1.2-$\alpha$1.8-H have been reached by ionising
photons and thus heated up, in the other three models many cells are
still cold and neutral.

There is also a significant amount of scatter in the temperature at
fixed density at all redshifts.  This scatter arises from the
different reionisation history of each cell in the simulation
(i.e. inhomogeneous reionisation) as well as the fact that we do not
use monochromatic photons, but rather a spectral energy distribution
which can also be hardened by spectral filtering
(\citealt{AbelHaehnelt99}).  This differs significantly from the
tight, power-law temperature-density relation expected in the
optically thin case following reionisation (\citealt{HuiGnedin97}).

There are also some small quantitative differences in the slope and
amplitude of the temperature-density relation
$T=T_{0}\Delta^{\gamma-1}$, which are summarised in
Table~\ref{table:tempdens}. It has been noted both observationally
(\citealt{Becker07}) and theoretically
(\citealt{Bolton04,Tittley07,Trac08,FurlanettoOh09}) that the
temperature-density relation may be multiple valued and inverted
following \HI reionisation.  This occurs because voids tend to be
reionised last and have therefore had less time to cool.  The
theoretical study of \cite{Trac08} in particular found $\gamma - 1\sim
-0.2$ at the end of reionisation.  These authors used a larger
simulation volume ($100h^{-1}$ Mpc) compared to this work, but found
the strong correlation between the density field and redshift of
reionisation in these models extends down to scales of $1h^{-1}$ Mpc.
We find the temperature-density relation is indeed very mildly
inverted ($\gamma-1 \sim -0.05$) for ${\mathcal E}$1.2-$\alpha$1.8-H
at $z=14$, but it remains close to isothermal for all other models at
all redshifts.  The origin of the diffferences between
  \cite{Trac08} and this work are not clear.  One possibility,
  however, is that \cite{Trac08} used a rather different prescription
  for the source emissivity based on the star formation implementation
  of \cite{Trac.Cen_2007}.  The ionising photon production rate in
  this model is not calibrated to match constraints from the \Lya
  forest data, and it therefore rises continuously toward lower
  redshift.  This means that the latter stages of reionisation occur
  more rapidly in their simulations compared to our model.  A more
  rapid end to reionisation could potentially explain the more
  strongly inverted temperature-density relation \cite{Trac08} find;
  proportionally more of the underdense gas will have been reionised
  and reheated close to the end of reionisation.

\begin{table*}
\caption{The temperature-density relation of the ionised IGM in our
  simulations. The columns indicate, from left to right, the name of
  the model, the redshift $z$, and the best fit power-law parameters
  for the power-law temperature-density relation
  $T=T_{0}\Delta^{\gamma-1}$. Here $T_0$ and $\gamma-1$ are calculated
  only in cells with $x_{\rm HII}>0.99$ (columns 3 and 5) and $x_{\rm
    HeIII}>0.99$ (columns 4 and 6).  See text for further details.}
\centering
\begin{tabular}{l | c cc c cc} 
\hline
Model & $z$ & \multicolumn{2}{c}{$T_0$ [K]} & \multicolumn{2}{c}{$\gamma$-1} \\
      &     & $x_{\rm HII}>0.99$ & $x_{\rm HeIII}>0.99$ & $x_{\rm HII}>0.99$ & $x_{\rm HeIII}>0.99$\\
\hline 
      & 14  & 16744   &   --      & -0.0404 &   --\\
      & 9   & 10525   &   --      &  0.0116 &   --   \\[-1ex]
\raisebox{1.5ex}{${\mathcal E}$1.2-$\alpha$1.8-H}  
      & 7   & 9293    &    --     &  0.0313 &  --     \\
      & 6   & 8648    &    --     &  0.0419 &   --    \\[1ex]
\hline
      & 14  & 19705   & 17027   & -0.0153 &  0.0101  \\
      & 9   & 14272   & 11735   &  0.0381 &  0.0567  \\[-1ex]
\raisebox{1.5ex}{${\mathcal E}$1.2-$\alpha$1.8}
      & 7   & 15823   & 14080   &  0.0367 &  0.0559  \\
      & 6   & 15927   & 15008   &  0.0341 &  0.0357  \\[1ex]
\hline
      & 14  & 18885   & 16743   & -0.0055 &  0.0123  \\
      & 9   & 14145   & 11796   &  0.0385 &  0.0641  \\[-1ex]
\raisebox{1.5ex}{${\mathcal E}$1.2-$\alpha$1-3}
      & 7   & 15826   & 12624   &  0.0370 &  0.0678  \\
      & 6   & 16236   & 14970   &  0.0351 &  0.0455  \\[1ex]
\hline
      & 14  & 14285   & 13828   &  0.0014 &  0.0179  \\
      & 9   & 10058   & 10753   &  0.0434 &  0.0438  \\[-1ex]
\raisebox{1.5ex}{${\mathcal E}$1.6-$\alpha$3}
      & 7   & 9922    & 11511   &  0.0554 &  0.0483  \\
      & 6   & 9725    & 13386   &  0.0594 &  0.0465  \\[1ex]
\hline
      & 14  & 14035   & 13779   &  0.0045 &  0.0169  \\
      & 9   & 10049   & 11005   &  0.0425 &  0.0412  \\[-1ex]
\raisebox{1.5ex}{${\mathcal E}$1.2-$\alpha$3}
      & 7   & 10649   & 11852   &  0.0437 &  0.0424  \\
      & 6   & 10468   & 13038   &  0.0453 &  0.0544  \\[1ex]

\hline 
\end{tabular}
\label{table:tempdens}
\end{table*}


\section{Implications for reionisation sources} \label{sec:observations}

In this section we now consider the implications our empirically
motivated simulations for reionisation by comparing them to
observational constraints on the IGM temperature at mean density, the
volume averaged neutral hydrogen fraction and recent estimates of the
ionising emissivity from measurements of the UV galaxy luminosity
function at $4<z<8$.

\subsection{The thermal state of the IGM at $z\simeq 5-6$}

\begin{figure}
\centering \includegraphics[width=0.55\textwidth]{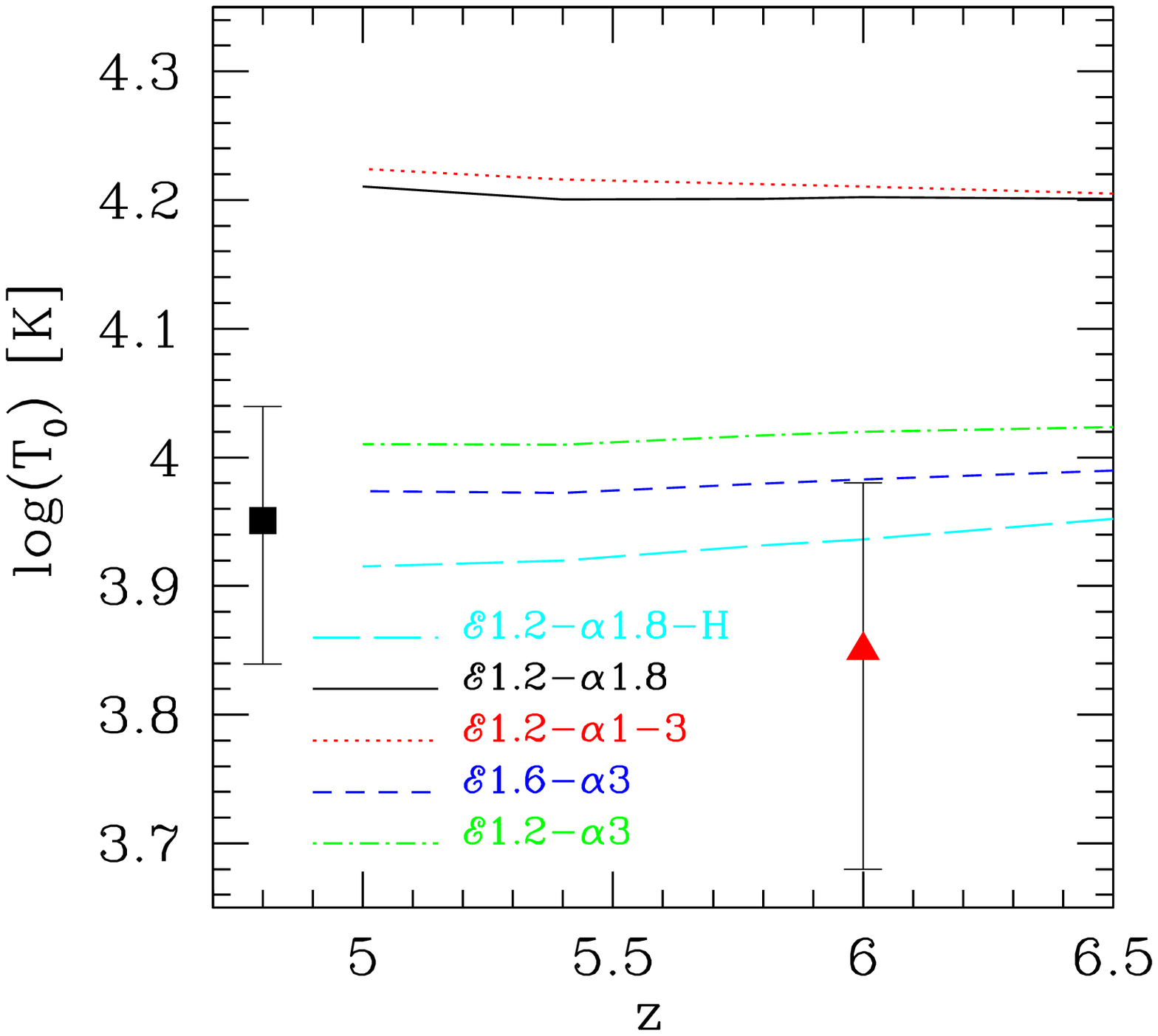}
\vspace{-1.3cm}
\caption{The IGM temperature at mean density, $T_0$, at different
  redshifts (see Table~\ref{table:tempdens}). The filled symbols
  refers to the values measured by \citet[][black square]{Becker11}
  and \citet[][red triangle]{Bolton12}.  The curves refer instead to
  the simulated results from models ${\mathcal E}$1.2-$\alpha$1.8-H
  (long dashed cyan), ${\mathcal E}$1.2-$\alpha$1.8 (solid black),
  ${\mathcal E}$1.2-$\alpha$1-3 (dotted red), ${\mathcal
    E}$1.6-$\alpha$3 (dashed blue) and ${\mathcal E}$1.2-$\alpha$3
  (dotted-dashed green).}
\label{fig:temp_obs}
\end{figure}

We first compare our simulations to recent measurements of the IGM
temperature in Figure~\ref{fig:temp_obs} (see also
\citealt{Raskutti12}). \citet{Becker11} recently presented constraints
on the thermal state of the IGM based on Ly$\alpha$ forest
observations in the redshift range $2.0 < z < 4.8$.  Their temperature
measurement at $z=4.8$ is reported as $T_{0} = 8930 \pm 2020$~K
($2\sigma$ errors) assuming an isothermal temperature-density relation
($\gamma=1$).  This constraint is shown by the black square in
Figure~\ref{fig:temp_obs}.  At higher redshift, $z \sim 6$,
\cite{Bolton12} have measured the temperature of the IGM within $\sim
5$ proper Mpc of seven quasars using the Doppler widths of Ly$\alpha$
absorption lines.  They report a line-of-sight averaged temperature at
the mean density of $T_{0}\sim 16200$~K.  Note, however, this
constraint is complicated by the fact that these quasars also reionise
the \HeII in their vicinity due to their hard ionising spectra.
\cite{Bolton12} therefore also provided an estimate for the
temperature after subtracting the expected heating from the local
reionisation of \HeII by the quasars, $T_{0}\sim 7100\rm\,K$, assuming
a quasar EUV spectral index of $\alpha=1.5$.  This latter estimate is
displayed in Figure~\ref{fig:temp_obs} as the red triangle with 95 per
cent confidence error bars.  Lastly, note that this constraint is
dependent on the uncertain amount of \HeII heating expected from the
quasars; assuming a harder (softer) EUV spectral index for the quasars
would lower (raise) this temperature constraint by several thousand
degrees.

Keeping this in mind, the curves in Figure~\ref{fig:temp_obs} display
the temperature at mean density, $T_0$, calculated in cells with
$x_{\rm HII}>0.99$ (see Table~\ref{table:tempdens}) in models
${\mathcal E}$1.2-$\alpha$1.8-H (long dashed cyan), ${\mathcal
  E}$1.2-$\alpha$1.8 (solid black), ${\mathcal E}$1.2-$\alpha$1-3
(dotted red), ${\mathcal E}$1.6-$\alpha$3 (dashed blue) and ${\mathcal
  E}$1.2-$\alpha$3 (dotted-dashed green).  We estimate the temperature
from the simulations in this manner to ensure any neutral gas which
has yet to be ionised is excluded; the temperature measurements from
the \Lya absorption measurements only probe highly ionised hydrogen.
The simulations which have a soft ($\alpha=3$) EUV spectral index
(${\mathcal E}$1.2-$\alpha$3 and ${\mathcal E}$1.6-$\alpha$3) as well
as the model which excludes helium (${\mathcal E}$1.2-$\alpha$1.8-H)
are similar or slightly greater than (within $\sim 0.02$ dex of the 95
per cent confidence interval) the measurement obtained by
\cite{Bolton12} at $z\sim 6$.  In contrast, the two models with harder
spectra (${\mathcal E}$1.2-$\alpha$1.8 and ${\mathcal
  E}$1.2-$\alpha$1-3) exhibit significantly higher temperatures due to
additional \HeII photo-heating.  Similarly, the \cite{Becker11}
temperature measurement at $z=4.8$ is also much lower than the
predicted simulation temperatures at $z=5$ for the harder ionising
spectra. Note that the agreement would be even worse if the
  heating contribution from X-rays were included in the simulations.

These results are thus consistent with a predominance of sources with
relatively soft ($\alpha \geq 3$) ionising spectra during hydrogen
reionisation, and also with an epoch of \HeII reionisation (most
likely driven by quasars) which was not fully underway until lower
redshift (e.g. \citealt{McQuinn2009}).  We therefore conclude that if
a population of sources with rather hard spectra, such as mini-quasars
(\citealt{Madau04}) or population-III stars (\citealt{Bromm01}) were
responsible for reionising hydrogen, their contribution must be either
{\it (i)} sub-dominant at all redshifts or {\it (ii)} confined
predominantly at early times ($z\geq 9$), such that there has been
sufficient time for the IGM temperature to cool and doubly ionised
helium to recombine by $z\simeq 6$.  This is not surprising as
  population-III stars are believed to be present at $z<9$, but,
  compared to population-II stars, in negligible numbers (see
  e.g. \citealt{Tornatore.Ferrara.Schneider_2007,Maio_etal_2010}). \cite{Becker12}
  have also recently pointed out that relative metal abundances in the
  IGM suggest population-II stars produced the bulk of hydrogen
  ionising photons during reionisation. Similarly, although
  mini-quasars have been investigated by a number of authors as
  possible sources of ionising photons, the general agreement is that
  their contribution is not dominant (see
  e.g. \citealt{Madau04,Miralda-Escude.Haehnelt.Rees_2000}).  In
  addition, a model in which reionisation were dominated by
  mini-quasars would most likely overpredict also the observed soft
  X-ray background \cite{Salvaterra.Haardt.Ferrara_2005}.

\subsection{Ionising photon production}  \label{sec:sources}                       

We next compare the ionising emissivity used in our simulations to
observational estimates based on recent measurements of the galaxy
luminosity function at $4<z<8$.  For this purpose, we compute the
ionising emissivity from galaxies using the recent fit to the redshift
evolution of the galaxy luminosity function presented by
\cite{Bouwens11}.  We assume a spectral energy distribution
$\epsilon_{\nu} \propto \nu^{0}$ for $912 {\rm \AA}<\lambda<3000 \rm
\AA$ and $\epsilon_{\nu} \propto \nu^{-3}$ (i.e. $\alpha=3$) for
$\lambda< 912 \rm \AA$, with an additional factor of six break at the
Lyman limit (e.g. \citealt{Leitherer99,Madau99}).  In addition, we
adopt two different redshift evolutions for the faint-end slope: the
\cite{Bouwens11} best fit $\alpha_{\rm LF} = -1.84 - 0.05(z-6)$, and a
steeper faint end slope of $\alpha_{\rm LF} = -1.9 - 0.1(z-6)$.  These
choices are intented to represent the considerable observational
uncertainty in the faint-end slope.

\begin{figure}
\centering \includegraphics[width=0.45 \textwidth]{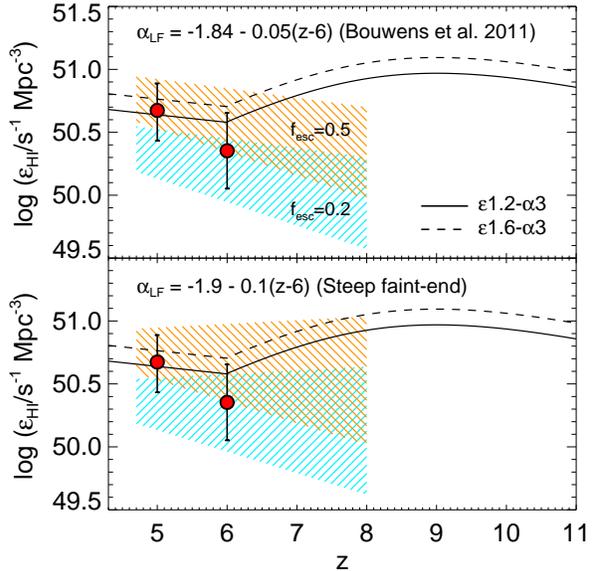}
\vspace{-0.3cm}
\caption{The ionising emissivity in models ${\mathcal E} 1.2-\alpha 3$
  (solid curve) and ${\mathcal E} 1.6-\alpha 3$ (dashed curve)
  compared to observational constraints based on \Lya forest data at
  $z \leq 6$ (red circles) and estimates of the emissivity from recent
  constraints on the luminosity function of high redshift Lyman break
  galaxies by \citet{Bouwens11} (hatched regions).  {\it Upper panel:}
  Comparison to the emissivity derived from the best fit redshift
  evolution of the luminosity function at $4<z<8$ presented by
  \citet{Bouwens11} (see text for details) with a faint end slope
  $\alpha_{\rm F}=-1.84 - 0.05(z-6)$.  The cyan and orange hatching
  assume escape fractions of $f_{\rm esc}=0.2$ and $f_{\rm esc}=0.5$,
  respectively, while the range of the hatched regions corresponds to
  the emissivity obtained by integrating the luminosity function to a
  lower magnitude limit of $M_{\rm lim}=-10$ and $M_{\rm lim}=-18$
  (upper and lower limit to hatching, respectively).  {\it Lower
    panel:} As for upper panel, but now assuming a steeper faint end
  slope for the luminosity function, $\alpha_{\rm LF} = -1.9 -
  0.1(z-6)$.}
\label{fig:emissivity}
\end{figure}

The resulting emissivities are displayed as the hatched regions in
Figure~\ref{fig:emissivity}, with the results from the two different
faint end slope evolutions shown in each panel.  The cyan and orange
hatching assume ionising photon escape fractions of $f_{\rm esc}=0.2$
and $f_{\rm esc}=0.5$, while the lower and upper limits to the
hatching correspond to the emissivity obtained by integrating the
\cite{Bouwens11} luminosity function fit to a lower magnitude limit of
$M_{\rm UV}=-18$ and $M_{\rm UV}=-10$, respectively.  These limits
roughly correspond to the magnitude limit of the observational data
and the expected magnitude of a galaxy in a halo with virial
temperature $2\times 10^{4}\rm\,K$ (\citealt{Trenti10}), respectively.
These are compared to the emissivities used in models ${\mathcal
  E}$1.2-$\alpha$3 (solid curve) and ${\mathcal E}$1.6-$\alpha$3
(dashed curve).  Observational constraints on the emissivity at $z\leq
6$ (red circles with error bars) derived from measurements of the
photo-ionisation rate from the \Lya forest
(\citealt{Wyithe.Bolton_2011}) and mean free path
(\citealt{SongailaCowie2010}) are displayed as red circles with error
bars.  Note again, that the models are by construction chosen to match
these constraints closely.

In order to match the emissivity in model ${\mathcal E}$1.2-$\alpha$3
up to $z=8$, an extrapolation of the faint end of the luminosity
function to $M_{\rm UV}=-10$, a high escape fraction $f_{\rm esc}=0.5$
and a slightly steeper faint-end slope than the best fit of
\cite{Bouwens11} are required.  Faint (and currently undetected)
galaxies are thus required to reproduce the ionising emissivity in our
simulations.  Recent theoretical studies indicate the faint end slope
may indeed steepen at $z>6$ (\citealt{Trenti10,Jaacks11}).  A rather
high Lyman continnum escape fraction is also required from these faint
galaxies.  Although impossible to measure directly at $z>6$, recent
observations indicate the escape fraction at $z\sim 3$ is larger than
at later times (e.g. \citealt{Siana10}).  In addition, \cite{Rauch11}
have recently presented observations of a morphologically disturbed,
faint \Lya emitting galaxy at $z=3.44$ which are consistent with a
Lyman continuum escape fraction of 50 per cent. These authors note
that such faint, interacting galaxies may be more common at higher
redshift, where the increasing importance of gravitational
interactions and mergers could provide a plausible mechanism for such
high escape fractions.

Finally, the emissivity evolution in our simulations is such that a
halo with a baryon mass $M_{\rm b}=10^8$~M$_\odot$ at $z=14$ produces
$\sim 5 \times 10^{52}$~phot~s$^{-1}$ and $\sim 10^{50}$~phot~s$^{-1}$
at $z=6$.  For comparison, the number of ionising photons emitted by a
halo with baryon mass $M_{\rm b} = M_{\rm tot}(\Omega_{\rm
  b}/\Omega_{\rm m})$ can be written as (see \citealt{Iliev_etal_2006a}):
\begin{eqnarray} 
\label{eq:phot}
\dot{N} & =  & \frac{f_{\star} f_{\rm esc} N_{\rm phot} M_{\rm b}}{m_{\rm p} \Delta t}  \\ \nonumber
& \simeq  & 5 \times 10^{52} {\,\rm phot \; s^{-1}}  \\ \nonumber
        & \times & \left(\frac{f_\star}{0.05}\right)  \left(\frac{f_{\rm esc}}{0.5}\right)
   \left(\frac{N_{\rm phot}}{5 \times 10^3}\right) \left(\frac{M_{\rm b}}{10^8 \; {\rm M_\odot}}\right)
   \left(\frac{10^7 \; {\rm yr}}{\Delta t}\right),
\end{eqnarray}

\noindent
where $f_\star$ is the fraction of baryons which are converted into
stars, $f_{\rm esc}$ is the escape fraction of ionising photons,
$N_{\rm phot}$ is the number of ionising photons per stellar baryon,
$m_{\rm p}$ is the proton mass and $\Delta t$ is the time between
  two snapshots of the hydrodynamical simulation\footnote{Note that
    the physically relevant timescale here is actually the lifetime of
    the stellar population. In practice, however, numerical
    simulations assume a uniform emission of ionising photons within
    each $\Delta t$, so that the total number of emitted photons is
    conserved.  For a more extensive discussion on Eq.~\ref{eq:phot}
    we refer the reader to the original paper.}.  Typically, $N_{\rm
  phot}=5 \times 10^3$ and $1\times 10^4$ for population-II stars with
a Salpeter IMF and a top-heavy IMF, respectively
(e.g. \citealt{Iliev_etal_2006a}).  The requirement for a large escape
fraction ($f_{\rm esc}\sim 0.5$) may be therefore relaxed somewhat if
the efficiency of ionising photon production increases toward higher
redshift or a top-heavy IMF is invoked (see
e.g. \citealt{Bromm.Ferrara.Coppi.Larson_2001,Schneider.Ferrara.Natarajan.Omukai_2002}).
However, as noted in the previous section, the IGM temperature
measurements appear to rule out significant reionisation by metal-free
stellar populations, at least at $z<9$.  However, as there are a
variety of possible parameter combinations which could satisfy the
emissivity required, it is not possible to set a stringent constraint
on the individual parameters in Eq.~(\ref{eq:phot}).

\subsection{The volume averaged \HI fraction}

Lastly, we compare our simulations to constraints on the volume
averaged \HI fraction, $x_{\rm HI}$, in the IGM at $z \geq 6$.  As
discussed earlier, the presently available observational data remain
inconclusive with regard to the redshift evolution of $x_{\rm HI}$.
This is largely because almost all the methods used to derive $x_{\rm
  HI}$ are somewhat model dependent and/or are limited by the
available data.  For example, at $z=5.5$, studies of the transmitted
flux in the \Lya forest indicate $x_{\rm HI} \sim 10^{-4}$
\citep{Fan06,Becker07,BoltonHaehnelt07} in the regions where \Lya
transmission is detected.  However, \cite{Mesinger10} has noted that
the relatively small number of quasar sight-lines which have been
analysed, combined with the fact that quasars sit in highly biased
regions, does not preclude an IGM which is still a few per cent
neutral by volume at $z=5$--$6$; isolated patches of neutral hydrogen
may still lurk undetected in the diffuse IGM at these redshifts due to
the inhomogeneous nature of reionisation (see also \citealt{Lidz07}).
Indeed, taking an (almost) model independent approach,
\cite{McGreer11} calculated a conservative upper limit of $x_{\rm HI}
\la 0.9$ from \Lya forest data at $z \sim 6.1$, although a subsample
of two deep spectra provided a more stringent constraint of $x_{\rm
  HI} \la 0.5$.

Alternative analyses of higher redshift quasar spectra also provide
variable estimates.  An analysis of a putative IGM damping wing in a
quasar near-zone at $z=6.28$ by \cite{MesingerHaiman04} yields $x_{\rm
  HI} \ga 0.2$.  In contrast,
\citet{Maselli.Gallerani.Ferrara.Choudhury_2007} find that the sizes
of quasar near-zones are consistent with an IGM which is mostly
ionized at $z\simeq 6$, with $x_{\rm HI} \la 0.06$.  More recently, an
analysis of the near-zone in the spectrum of the highest redshift
quasar yet detected was found to be consistent with $x_{\rm HI} \ga
0.1$ at $z=7.085$ \citep{Mortlock11,Bolton11}.  However, all of these
observations probe only the neutral fraction in the vicinity of these
quasars, so the interpretation of these measurements with respect to
the IGM as a whole is again hampered by the inhomogeneous nature of
reionisation (\citealt{MesingerFurlanetto08}).  Lastly, recent
measurements of a rapid decline in the \Lya emitter/Lyman break galaxy
fraction indicate the neutral fraction may be as high as $x_{\rm HI}
\sim 0.5$ at $z\sim 7$ (\citealt{Schenker11,Pentericci11,Ono11}).  On
the other hand, the effect of patchy reionisation and galactic
outflows on reionisation also complicate the use of \Lya emitting
galaxies as a probe of the volume averaged neutral fraction
(e.g. \citealt{Dijkstra11}).

\begin{figure}
\centering \includegraphics[width=0.55\textwidth]{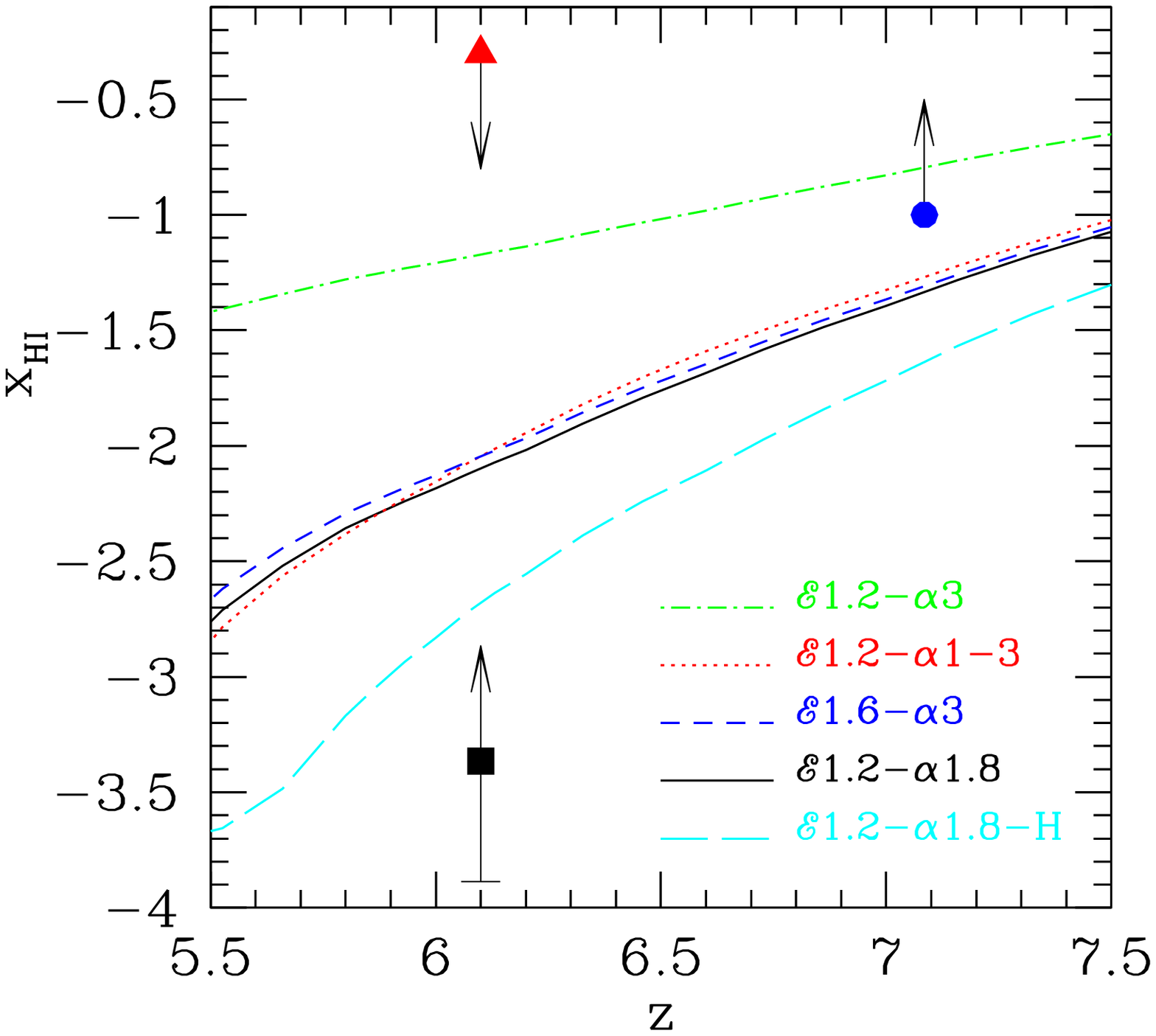}
\vspace{-1.3cm}
\caption{The volume averaged \HI fraction and its evolution with
  redshift. The filled symbols refer to the observational measurements
  by \citet[][black square]{Fan06}, \citet[][red triangle]{McGreer11}
  and \citet[][blue circle, see text for further
    details]{Mortlock11}. The curves display the results of our
  radiative transfer simulations: ${\mathcal E}$1.2-$\alpha$1.8-H
  (long dashed cyan), ${\mathcal E}$1.2-$\alpha$1.8 (solid black),
  ${\mathcal E}$1.2-$\alpha$1-3 (dotted red), ${\mathcal
    E}$1.6-$\alpha$3 (dashed blue) and ${\mathcal E}$1.2-$\alpha$3
  (dotted-dashed green).}
\label{fig:xHI_obs}
\end{figure}

In Figure~\ref{fig:xHI_obs} we present a comparison between the volume
averaged neutral fraction predicted by our simulations and a selection
of these measurements.  There are two important points to note here.
Firstly, all our simulations lie within the
(admittedly large) region between the lower and upper limits at
$z\simeq 6$.  However, the \cite{Mortlock11} measurement appears to
exclude all the models with the exception of ${\mathcal
  E}$1.2-$\alpha$3; the neutral fraction in all the other cases is too
low and as a consequence the emissivity is too high.  Reconciling
these models with the \cite{Mortlock11} neutral fraction at $z\sim
7.1$ would therefore require a lower ionising emissivity which then
must remain constant or even increase weakly toward lower redshift to
simultaneously match the $z=6$ photo-ionisation rate measurements.  On
the other hand, \cite{Bolton11} note that uncertainties in the
abundance of high column density systems and the spectral shape of the
quasar ionising radiation could weaken the upper limit on $x_{\rm
  HI}$, so the significance of this difference should be treated
cautiously.

The second (related) point is that all four models which include
helium predict a neutral fraction at $z=6$ between $1$--$6$ per cent,
which lies 1--2 orders of magnitude {\it above} the constraints from
the \Lya forest opacity.  This is in stark contrast to the
conventional interpretation that the IGM is highly ionised, $x_{\rm
  HI}\sim 10^{-4}$, by $z=6$, although this scenario is consistent
with the conservative estimates of \cite{McGreer11}.  This result is
perhaps not too surprising; numerical models which predict a highly
ionised IGM at $z=6$ typically overpredict the photo-ionisation rate
or ionising intensity by a factor of two or more
(e.g. \citealt{Iliev08,Finlator09b,Aubert10}).  This implies that when
we deliberately match the emissivity in our simulations at $z=6$ to the
\Lya forest data, the IGM is required to have an appreciable neutral
fraction at $z\sim 6$.  A more highly ionised IGM by $z=6$ may be
obtained by adopting an ionising emissivity which increases more
rapidly than we already assume at $z>6$, but this would still come at
the expense of not satisfying the $z\sim 7$ neutral fraction
constraint.

An important caveat, however, is that most reionisation models
(including this work) do not correctly resolve Lyman limit systems
(although see \citealt{Kohler07,McQuinn11}).  Lyman limit systems
(LLSs) are expected to regulate the mean free path of ionising photons
once the sizes of ionised bubbles exceed the typical separation
between these optically thick systems
(\citealt{GnedinFan06,FurlanettoMesinger09}).  Since the \HI
photo-ionisation rate is proportional to the emissivity and the mean
free path, $\Gamma_{\rm HI} \propto \epsilon_{\rm HI}\lambda_{\rm
  HI}$, correctly modelling LLSs is a crucial ingredient for
simulating the latter stages of reionisation.  Although our
simulations match the observational measurements of $\Gamma_{\rm HI}$
by design, the mean free path within the simulations is not set by
LLSs, but rather the remaining patches of neutral gas in the IGM which
are furthest from the ionising sources (in the case of ${\mathcal
  E}$1.2-$\alpha$3, this is 6 per cent of the IGM by volume at $z=6$).
A mean free path at $z=6$ which is instead set by LLSs might allow for
an emissivity which is consistent with the observed constraints on
$\Gamma_{\rm HI}$, but at the same time have a lower volume averaged
neutral fraction due to the smaller volume filling factor of these
dense optically thick systems.

Note again, however, that the issue of how one could then reconcile
the large volume averaged neutral fraction of $x_{\rm HI}>0.1$ at
$z=7.1$ with {\it (i)} a low neutral fraction of $x_{\rm HI}\sim
10^{-4}$ at $z=6$ and {\it (ii)} an emissivity at $z=6$ equivalent to
$\sim 1$--$3$ ionising photons emitted per hydrogen atom over a Hubble
time remains.  Since the emissivity must increase at $z>6$ for
reionisation to complete {\it by} $z=6$ (\citealt{BoltonHaehnelt07}),
either the IGM is more highly ionised at $z\sim 7$ than recent
observations suggest, or the IGM is still a few per cent neutral by
volume at $z=6$ (\citealt{Mesinger10}).


\section{Summary and conclusions}

In this work we have investigated the impact of helium on hydrogen
reionisation using three dimensional, multi-frequency RT simulations.
We performed five simulations using different models for the amplitude
and spectral shape of the ionising emissivity during reionisation.  By
design, all our models are consistent with measurements of the Thomson
scattering optical depth and the metagalactic hydrogen
photo-ionisation rate at $z \sim 6$.  This empirical approach enables
us to explore the consequences of satisfying these observational
constraints for reionisation.  The main outcomes of this study may be
summarised as follows.

\begin{itemize}
\item The evolution of the volume averaged \HII fraction, $x_{\rm
  HII}$, is very similar for all models with the same hydrogen
  ionising emissivity independent of the EUV spectral index.  However,
  the spectral energy distribution has a strong impact on the
  evolution on the volume averaged \HeII and \HeIII fractions, $x_{\rm
    HeII}$ and $x_{\rm HeIII}$. Models with a soft power-law EUV
  index, $\alpha=3$, produce a much lower $x_{\rm HeIII}$ compared to
  models in which harder photons are present. The inclusion of helium
  in the RT simulations furthermore slightly delays reionisation due
  to the small number of ionising photons which reionise neutral
  helium instead of hydrogen.\\

\item The choice of EUV spectral index has a significant effect on the
  evolution of the volume averaged IGM temperature during
  reionisation.  At $z \simgt 10$, model ${\mathcal
    E}$1.2-$\alpha$1.8-H (without helium) has a volume averaged
  temperature which is $\sim 10$ per cent higher than the
  corresponding model including helium, ${\mathcal E}$1.2-$\alpha$1.8,
  due to the slightly larger volume of the IGM which is photo-ionised
  by this time.  However, at lower redshift the inclusion of \HeII
  photo-ionisation results in a higher volume averaged temperature for
  ${\mathcal E}$1.2-$\alpha$1.8.  In comparison, despite exhibiting
  behaviour similar to ${\mathcal E}$1.2-$\alpha$1.8 and ${\mathcal
    E}$1.2-$\alpha$1-3 for the evolution of the \HII filling factor,
  the softer ionising spectrum used in ${\mathcal E}$1.6-$\alpha$3
  produces volume averaged temperatures which are 20-30 per cent lower
  than ${\mathcal E}$1.2-$\alpha$1.8.  This is partly because the
  volume filling factor of \HeIII is smaller in this model, but also
  because the softer ionising photons produce less photo-heating.\\

\item The temperature (and ionisation fraction) distributions in the
  simulations exhibit a significant amount of scatter at all
  redshifts.  This scatter arises from the different reionisation
  history of each cell in the simulations (i.e. inhomogeneous
  reionisation) as well as the fact that we do not use monochromatic
  photons, but rather a spectral energy distribution which can also be
  hardened by spectral filtering.  This differs significantly from the
  tight, power-law temperature-density relation expected for an
  optically thin IGM following reionisation.  We find the
  temperature-density relation for ionised gas is typically isothermal
  or mildly inverted during hydrogen reionisation. \\

\item A comparison with recent estimates of the IGM temperature at $z
  \sim 5-6$ from \Lya absorption in the spectra of high redshift
  quasars suggests that hydrogen reionisation is mainly driven by
  sources with a soft spectral energy distribution, $\alpha \leq 3$.
  The simulations with harder spectral indices produce temperatures
  which are larger than the observational constraints.  We conclude
  that population-II stellar sources are likely to provide most of the
  ionising photons during reionisation, and the spectral shape of the
  ionising background must harden at $z<6$ due to the increasing
  importance of quasars if \HeII reionisation is to complete by
  $z\simeq 3$.  If sources with rather hard spectra, such as
  mini-quasars or population-III stars were responsible for reionising
  hydrogen, their contribution must be either small or confined to
  $z\geq 9$ to give sufficient time for the IGM temperature to cool
  and for doubly ionised helium to recombine by $z\simeq 6$.\\

\item In order to reproduce the ionising emissivity in our simulations
  at $z>6$, we find that the best fit to the evolution of the galaxy
  luminosity function presented by \cite{Bouwens11} at $4<z<8$
  requires extrapolation to faint UV magnitudes ($M_{\rm UV} =-10$),
  as well as a steepening faint end slope $\alpha_{\rm LF} \leq -2$
  and a high Lyman continuum escape fraction $f_{\rm esc}=0.5$.
  Faint, low mass galaxies are therefore necessary for providing the
  required number of photons during reionisation, in agreement with
  several other complementary studies.\\

\item There is some tension between the empirically motivated ionising
  emissivity used in our simulations and recent observational
  constraints on the IGM neutral fraction which indicate that $x_{\rm
    HI}>0.1$ at $z\sim 7.1$.  The ionising emissivity inferred from
  the \Lya forest at $z=6$ is equivalent to only $1$--$3$ ionising
  photons emitted per hydrogen atom over a Hubble time, implying
  reionisation is extended and that the emissivity must increase at
  $z>6$ if reionisation is to complete by $z=6$
  (\citealt{MiraldaEscude2003,BoltonHaehnelt07}).  However, an
  increasing emissivity at $z>6$ is inconsistent with a large neutral
  fraction at $z\sim 7$ in our simulations unless the observations are
  overestimates or the IGM remains a few per cent neutral by volume at
  $z=6$ (see e.g \citealt{Mesinger10}).

\end{itemize}

\noindent
Our results highlight the importance of reproducing post-reionisation
constraints such as the IGM temperature and background
photo-ionisation rate for constraining reionisation models.  While
these simulations were designed mainly to investigate the impact of
helium on hydrogen reionisation and the sources of ionising photons at
high redshift, the volume used is too small to allow a more detailed
discussion on helium reionisation (which is thought to be driven by
quasars and to be complete at $z \sim 2.5-3$) and a more accurate
comparison with observational constraints at $z<6$.  We will postpone
this further analysis to a future work, together with a more thorough
investigation of the impact of unresolved small scale high density
peaks.  The latter will be particularly important for regulating the
tail-end of the reionisation process and for setting the thermal state
of the IGM by absorbing photons close to the \HI and \HeII ionisation
edges.  Including these effects in numerical models is therefore
necessary for refining the comparison of simulations with
observations at $z<6$.

\section*{acknowledgments}

The authors would like to thank an anonymous referee for his/her very
constructive comments, and K. Finlator and A. Meiksin for useful
suggestions.  The hydrodynamical simulation used in this work was
performed using the Darwin Supercomputer of the University of
Cambridge High Performance Computing Service
(http://www.hpc.cam.ac.uk/), provided by Dell Inc. using Strategic
Research Infrastructure Funding from the Higher Education Funding
Council for England. BC acknowledges the hospitality of the 4C
Institute at the Scuola Normale Superiore of Pisa.  JSB acknowledges
the support of an ARC postdoctoral fellowship (DP0984947).  AM
acknowledges the support of the DFG Priority Program 1177.

\bibliographystyle{apj}
\bibliography{helium.bib}

\begin{appendix}

\section{Convergence tests}

\begin{figure*}
\includegraphics[width=1.00\textwidth]{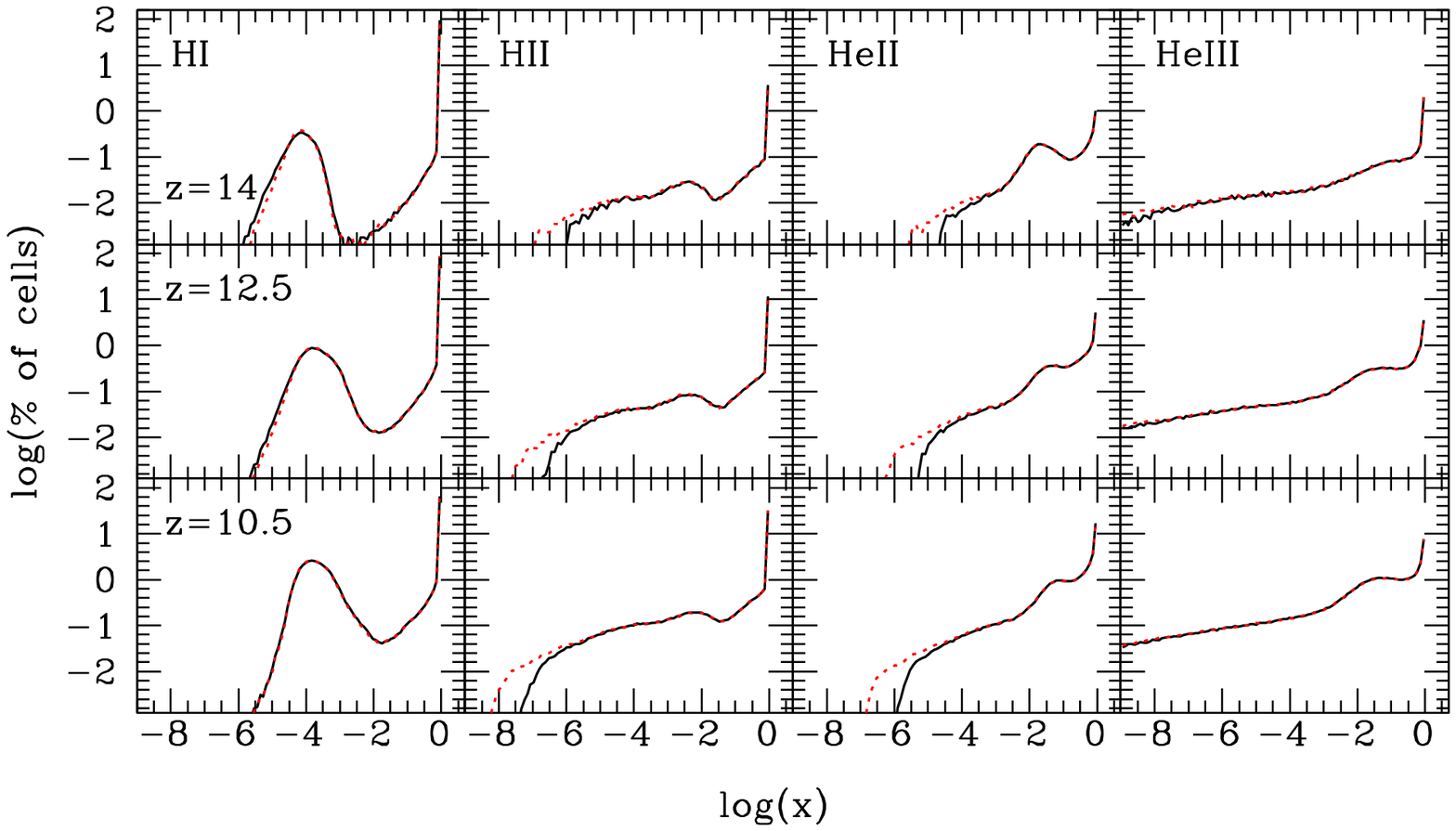}
\vspace{-1cm}
\caption{The percentage of cells in the radiative transfer simulations
  as a function of H$\,\rm \scriptstyle I$, H$\,\rm \scriptstyle II$,
  He$\,\rm \scriptstyle II$ and He$\,\rm \scriptstyle III$ fractions
  (from left to right) at $z$=14 (upper row), 12.5 (middle row) and 10.5
  (lower row).  The curves in each panel correspond to model
  ${\mathcal E}$1.2-$\alpha$1.8 (solid black lines) and the same model run with
  10 times more photon packets (red dotted).} 
\label{fig:distr_x_conv}\end{figure*}

\begin{figure}
\includegraphics[width=0.40\textwidth]{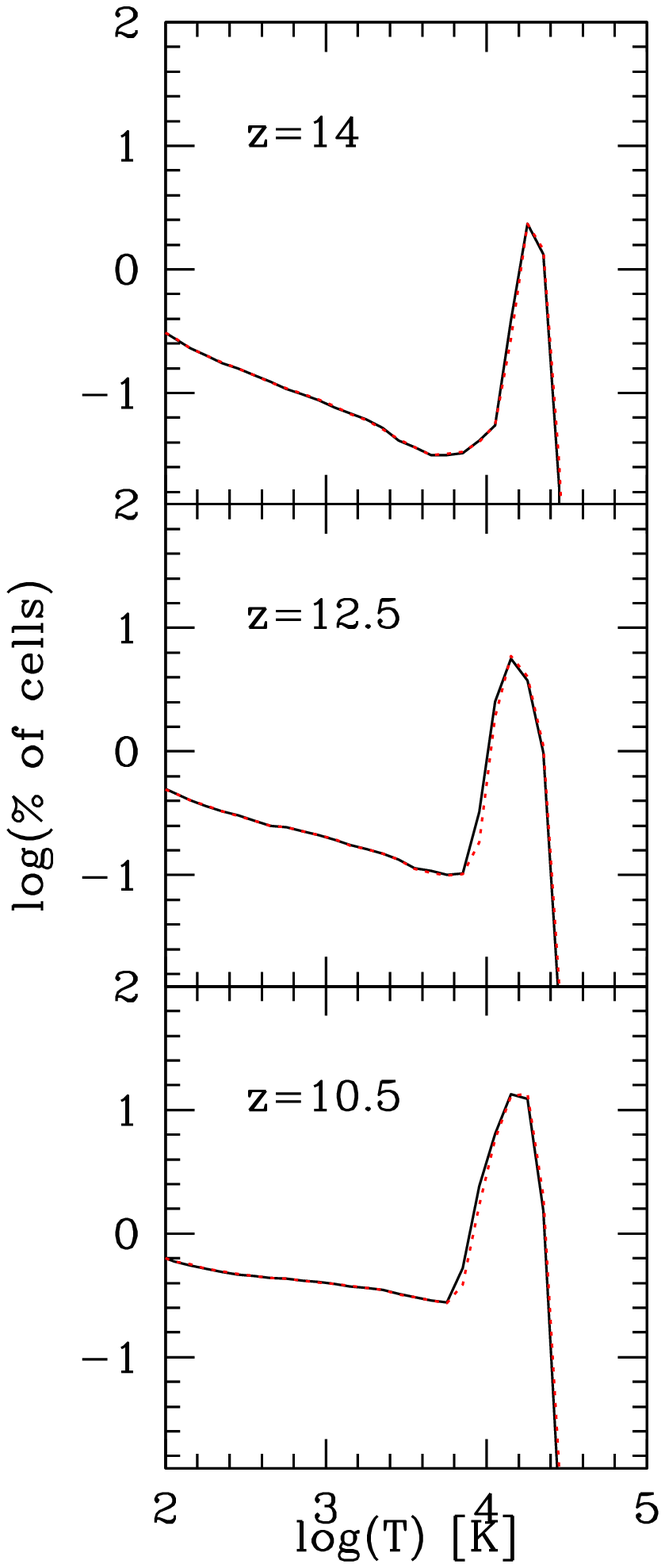}
\vspace{-1cm}
\caption{The percentage of cells in the radiative transfer simulations
  as a function of the gas temperature $T$ at $z$=14 (upper row), 12.5 (middle row) and 10.5
  (lower row).  The curves in each panel correspond to model
  ${\mathcal E}$1.2-$\alpha$1.8 (solid black lines) and the same model run with
  10 times more photon packets (red dotted).} 
\label{fig:distr_t_conv}\end{figure}

As discussed in Section~\ref{radtrans}, depending on the redshift and number of sources, 
we emit $10^5-10^6$ photon packets per source at each $t_{rt,i}$, corresponding to 
a total of $\sim 5 \times 10^7-10^{10}$ photon packets. While it is computationally
too expensive to run a full simulation with an order of magnitude more photon packets, 
we have run tests on single snapshots and on a limited number of consecutive snapshots 
at high redshift.  In Figures~\ref{fig:distr_x_conv} and~\ref{fig:distr_t_conv} the distribution of 
different species and gas temperature, respectively, is shown for run 
${\mathcal E}$1.2-$\alpha$1.8 (black solid lines) and for the same simulation with 10
times more photon packets (red dotted).  The results are shown down to the lowest redshift reached by the higher resolution simulation, i.e. $z=10.5$, which is obtained using 12 snapshots of the hydrodynamic simulation.  It is evident that an excellent convergence has 
been reached both for the H and He species and the gas temperature, with
the exception of cells with $x_{\rm HII}<10^{-6}$ and $x_{\rm HeII}<10^{-4}$.
Tests using only one snapshot at lower redshifts (i.e. following the radiative transfer starting from a non neutral configuration) show a similar convergence, but they do not account for differences between the two runs which might have accumulated if the full reionisation history were followed.
The above Figures though demonstrate that such differences are negligible.

\end{appendix}

\label{lastpage}
\end{document}